\documentclass[11pt,psfig]{article}
\usepackage{amsmath}
\usepackage{amsfonts}
\usepackage{graphics}
\usepackage{graphicx}
\usepackage{ragged2e}
\usepackage{float}
\usepackage{latexsym}
\usepackage{amsthm}
\usepackage{cite}
\usepackage{amssymb}
\usecounter{enumi}
\newtheorem{Lemma}{Lemma}[section]
\newtheorem{Theorem}[Lemma]{Theorem}

\newtheorem{Corollary}[Lemma]{Corollary}
\newtheorem*{Proof}{Proof}
\newtheorem{Proposition}[Lemma]{Proposition}

\numberwithin{equation}{section}

\def\be{\begin{eqnarray}} \def\ee{\end{eqnarray}} 
  \def\({\left(} \def\){\right)}
\def\bc{\begin{center}} 
\def\ec{\end{center}}  

\def\bey{\begin{eqnarray*}}\def\eey{\end{eqnarray*}}
\begin{document}
\title{Local isometric immersions of pseudospherical surfaces described by a class of third order partial differential equations}

\author{ Mingyue Guo$^{a}$, Zhenhua Shi$^{a,b}$ \footnote{Corresponding author, E-mail address:
zhenhuashi@nwu.edu.cn}
\vspace{4mm}\\
$^{a}$\small School of Mathematics, Northwest University, Xi'an 710069,  P.R. China\\
$^{b}$\small Center for Nonlinear Studies, Northwest University, Xi'an 710069,  P.R. China}
\date{}
\maketitle
 \bc
\begin{minipage}{130mm}
{\bf Abstract}\\ In this paper, we study the problem of local isometric immersion of pseudospherical surfaces determined by the solutions of a class of third order nonlinear partial differential equations with the type $u_t - u_{xxt} = \lambda u^2 u_{xxx} + G(u, u_x, u_{xx}),(\lambda\in\mathbb{R})$. We prove that there is only two subclasses of equations admitting a local isometric immersion into the three dimensional Euclidean space $\mathbb{E}^3$ for which the coefficients of the second fundamental form depend on a jet of finite order of $u$, and furthermore, these coefficients are universal, namely, they are functions of $x$ and $t$, independent of $u$. Finally, we show that the generalized Camassa-Holm equation describing pseudospherical surfaces has a universal second fundamental form.

\end{minipage}
\ec

\section{Introduction}

\indent \indent The study of partial differential equations describing pseudospherical surfaces holds significant importance, as these equations possess both theoretical importance and practical relevance in mathematics and physics. Chern and Tenenblat introduced the concept of a partial differential equation which describes pseudospherical surfaces in \cite{Chern1986}, inspired by Sasaki's observation \cite{Sasaki1979}. Their remarkable works established connections between solutions of partial differential equations and differentiable two-dimension Riemannian manifolds, and profoundly influenced a succession of subsequent papers \cite{Jorge1987,Rabelo1989,Reyes1998,Ding2002,Silva2015,Neto2010,Ferraioli2020,Kelmer2022,Ferraioli2024,Kelmer2025} and etc..

A $k$-th order partial differential equation for a real-valued function $u(x, t)$ is said to describe pseudospherical surfaces if there exists 1-forms $\omega_i = f_{i1} dx + f_{i2} dt$, $1 \leq i \leq 3$, where the coefficient functions $f_{ij}$, $j = 1, 2$, depend on $u(x, t)$ and its derivatives, such that the structure equations of a surface with Gaussian curvature $K = -1$, say
\begin{equation}\label{1.1}
	d\omega_1 = \omega_3 \wedge \omega_2, \quad d\omega_2 = \omega_1 \wedge \omega_3, \quad d\omega_3 = \omega_1 \wedge \omega_2,
\end{equation}
are satisfied whenever $u(x,t)$ is a solution of the partial differential equation for which
\begin{equation}\label{1.2}
\omega_{1} \wedge \omega_{2}\neq 0.
\end{equation}
The 1-form $\omega_3$, known as the Levi-Civita connection form, is uniquely defined by the 1-forms $\omega_1$ and $\omega_2$. Together, these three 1-forms determine the first fundamental form, given by
\begin{equation}\label{1.3}
I=\omega_1^2+\omega_2^2.
\end{equation}

The property of a surface being pseudospherical is intrinsic as it is entirely determined by its first fundamental form \cite{Kahouadji2015}. On the other hand, due to the well-known result that any pseudospherical surface can be locally isometrically immersed into $\mathbb{E}^3$, the problem of local isometric immersion of pseudospherical surfaces determined by the solutions of partial differential equations arises from an extrinsic perspective. In a recent series of papers \cite{Kahouadji2015,Kahouadji2016,Silva2016,Ferraioli2017,Kahouadji2019,Ferraioli2022}, researchers addressed the local isometric immersion problem of pseudospherical surfaces described by the equations studied in \cite{Chern1986,Rabelo1990,Silva2015, Ferraioli2016,Ferraioli2020}. For the purposes of this paper, the motivation comes from the special feature of the sine-Gordon (SG) equation
\begin{equation}\label{1.4}
u_{xt}=\sin u,
\end{equation}
which was initially identified as being equivalent to the Gauss-Codazzi equations for pseudospherical surfaces in classical surface theory using Darboux asymptotic coordinates \cite{Bour1862}. It is straightforward to prove that the following 1-forms
\begin{equation}\label{1.5}
\omega_1=\frac{1}{\eta}\sin u dt,\quad \omega_2=\eta dx+\frac{1}{\eta}\cos  u dt, \quad \omega_3=u_x dx,
\end{equation}
where $\eta$ is a non-vanishing real parameter, satisfy the structure equations (\ref{1.1}) whenever $u$ is a solution of the SG equation (\ref{1.4}), and thus, it describes pseudospherical surfaces. Moreover, its first and second fundamental forms are given by, respectively,
\begin{equation}\label{1.6}
I=\frac{1}{\eta^2}dt^2+2\cos u dxdt+\eta^2 dx^2, \quad II=\pm 2 \sin u dxdt.
\end{equation}
Observe that the coefficients of the second fundamental form of the SG equation (\ref{1.4}) depends on a jet of order zero of $u$. A natural question emerges: are there other partial differential equations describing pseudospherical surfaces, whose coefficients of the second fundamental form of the local isometric immersion depend on a jet of finite order of $u$, including $x$, $t$, $u$, and a limited number of $u$'s derivatives? Unexpectedly, the findings in \cite{Kahouadji2015,Kahouadji2016,Silva2016,Ferraioli2017,Kahouadji2019,Ferraioli2022,Freire2022} suggest that, with the exception of the SG equation and short pulse equation along with its some generalizations, most equations describing pseudospherical surfaces likely admit a local isometric immersions into $\mathbb{E}^3$ with universal second fundamental form, that is, the coefficients of the second fundamental form depend only on $x$ and $t$. These conclusions highlights the unique significance of the SG equation and short pulse equation and their generalizations among all equations describing pseudospherical surfaces. It also encourages further exploration to discover more examples which possess such remarkable property.

This paper is concerned with the dependence of the second fundamental form of  local isometric immersion on the solution $u$ of third order partial differential equations with the type
\begin{equation}\label{1.7}
u_t - u_{xxt} = \lambda u^2 u_{xxx} + G(u, u_x, u_{xx}), \quad \lambda \in \mathbb{R},
\end{equation}
which describe pseudospherical surfaces with associated 1-forms
\begin{equation}\label{1.8}
\omega_i = f_{i1}\,dx + f_{i2}\,dt, \quad 1 \leq i \leq 3,
\end{equation}
satisfying the auxiliary condition
\begin{equation}\label{1.9}
f_{p1} = \mu_p f_{11} + \eta_p, \quad \mu_p, \eta_p \in \mathbb{R}, \quad p = 2, 3.
\end{equation}
The classification of the equations (\ref{1.7}) was systematically established in our previous work, where we also demonstrated that a series of typical soliton equations belong to certain subclass, such as the generalized Camassa-Holm (CH) equation
\begin{equation}\label{1.10}
 u_t - u_{xxt} = u^2 u_{xxx} - u^2u_{xx}-3uu_x^2-2u^2u_x+4u u_x u_{xx}+u_x^3,
\end{equation}
which is one of the most well-known equations proposed by Novikov \cite{Novikov2009}.

Our main result is as follows:
\begin{Theorem}\label{theorem1.1}
Consider a class of third order partial differential equations of the form (\ref{1.7}) describing pseudospherical surfaces, only the following two families of equations
\begin{equation}\label{1.11}
u_t-u_{xxt}=\frac{1}{f'}\left(\phi_{1,u}u_{x}+\phi_{1,u_x}u_{xx}\pm \frac{\eta_2}{\sqrt{1+\mu_{2}^2}}\phi_{1}\right),\quad \eta_2\phi_{1}\neq0,
\end{equation}
and
\begin{equation}\label{1.12}
\begin{split}
u_t - u_{xxt}=& \lambda u^2 u_{xxx} + \frac{1}{f'} \left[ u_x \phi_{1,u} + u_{xx} \phi_{1,u_x} - \lambda u^2 u_x f' \pm \frac{\eta_2}{\sqrt{1+\mu_{2}^2}}\phi_{1} \right. \\
& \left. -\left(2\lambda uu_x\pm\frac{\eta_2}{\sqrt{1+\mu_{2}^2}}\lambda u^2 \pm\frac{C_1}{\sqrt{1+\mu_{2}^2}}\right)f\right], \quad C_1 \in \mathbb{R},
\end{split}
\end{equation}
where $(\lambda \eta_2)^{2}+C_1^{2} \neq 0$, and $f=f(u-u_{xx})$ and $\phi_{1}=\phi_{1}(u,u_x)$ are differentiable functions with $f'\neq 0$,
exist a local isometric immersions into $\mathbb{E}^3$ of the surfaces determined by the solutions $u$, for which the coefficients of the second fundamental forms depend on a jet of finite order of $u$. Moreover, these coefficients are universal functions of $x$ and $t$, independent of $u$.
\end{Theorem}

\begin{Corollary}\label{corollary1.2}
The generalized CH equation (\ref{1.10}) describes pseudospherical surfaces and has a universal second fundamental form.
\end{Corollary}

The remainder of this paper is organized as follows: In Section 2, we give a brief review in some basic facts concerning the classical theory of pseudospherical surfaces and provide, without proof, the classification results from our previous work that are essential for proving Theorem 1.1. The detailed proof of the main theorem are presented in Section 3.

\section{Preliminaries}

\indent \indent In this section, we recall some fundamental facts from classical theory of pseudospherical surfaces. For further details, the interested reader is referred to \cite{Cartan1970,Ivey2003,Tenenblat1998}.

\subsection{Total derivatives and prolongations}

\indent \indent Hereafter, for the convenience of discussion, we adopt the notation
\begin{equation}\label{2.1}
 u_i=\partial_x^i u,\quad w_j=\partial_t^j u, \quad v_k=\partial_t^k u_x,\quad 0\leq i\leq l,0\leq j\leq m,0\leq k\leq n,
\end{equation}
where $u_0=w_0=u$ and $u_1=v_0=u_x$. Therefore, the total derivatives of a differential function  $h=h(x,t,u_0,u_1,\dots, u_l,w_1,\dots,w_m, v_1,\dots,v_n)$, $1\leq l,m,n\leq\infty$, are given by
\begin{equation}\label{2.2}
D_x h=h_x+\sum_{i=0}^l h_{u_i}u_{i+1}+\sum_{j=1}^m h_{w_j}w_{j,x}+\sum_{k=1}^n h_{v_k}v_{k,x},
\end{equation}
\begin{equation}\label{2.3}
D_t h=h_t+\sum_{i=2}^l h_{u_i}u_{i,t}+\sum_{j=0}^m h_{w_j}w_{j+1}+\sum_{k=0}^n h_{v_k}v_{k+1},
\end{equation}
and the prolongations of the partial differential equation (\ref{1.7}) are expressed as
\begin{equation}\label{2.4}
u_{2q,t}=u_{0,t}-\sum_{i=0}^{q-1}D_x^{2i}F,\quad u_{2q+1,t}=u_{1,t}-\sum_{i=0}^{q-1}D_x^{2i+1}F,
\end{equation}
where $q=1,2,\dots,F(u_0,u_1,u_2,u_3)=\lambda u_0^2u_3+G(u_0,u_1,u_2)$ and $D_x^0 F=F$.

\subsection{Theory of pseudospherical surfaces}

\indent \indent Let $S$ be a pseudospherical surface in three-dimension Euclidean space $\mathbb{E}^3$, and its parametric equation is $\textbf{r}=\textbf{r}(t,x)$. Consider an orthonormal coordinate grid on $S$, and let $\mathbf{e}_1$, $\mathbf{e}_2$ be the unit tangent vectors, and $\mathbf{e}_3$ be the unit normal vector. Then the coframes $\omega_i$ and $\omega_{ij}$ satisfy
\begin{equation}\label{2.5}
\begin{cases}
\begin{split}
d\mathbf{r} &= \omega_1 \mathbf{e}_1 + \omega_2 \mathbf{e}_2, \quad \omega_3 = 0, \\
d\mathbf{e}_1 &= \omega_{12} \mathbf{e}_2 + \omega_{13} \mathbf{e}_3, \\
d\mathbf{e}_2 &= \omega_{21} \mathbf{e}_1 + \omega_{23} \mathbf{e}_3, \\
d\mathbf{e}_3 &= \omega_{31} \mathbf{e}_1 + \omega_{32} \mathbf{e}_2,
\end{split}
\end{cases}
\end{equation}
with $\omega_{ij}+\omega_{ji}=0$, $i,j=1,2,3$. The first and second fundamental forms read
\begin{equation}\label{2.6}
I=d\textbf{r} \cdot d\textbf{r}=\omega_1^2+\omega_2^2,
\end{equation}
and
\begin{equation}\label{2.7}
II=-d\textbf{r}\cdot d\textbf{e}_3=\omega_1\cdot\omega_{13}+\omega_2\cdot\omega_{23}.
\end{equation}
Since $d^2\textbf{r}= d^2\textbf{e}_i= 0$, we have the Cartan's structure equations
\begin{equation}\label{2.8}
d\omega_1=\omega_{12}\wedge\omega_2,\quad d\omega_2=\omega_{1}\wedge\omega_{12},
\end{equation}
\begin{equation}\label{2.9}
\omega_1 \wedge \omega_{13}+\omega_{2}\wedge\omega_{23}=0,
\end{equation}
and
\begin{equation}\label{2.10}
d\omega_{12}=\omega_{13}\wedge\omega_{32},
\end{equation}
\begin{equation}\label{2.11}
d\omega_{13}=\omega_{12}\wedge\omega_{23},\quad d\omega_{23}=\omega_{21}\wedge\omega_{13},
\end{equation}
The equations (\ref{2.11}) are called the Codazzi equations in classical theory of surfaces. It follows from the equation (\ref{2.9}) that $\omega_1 \wedge \omega _2\wedge \omega_{13}=\omega_1 \wedge \omega _2\wedge\omega_{23}=0$ and by using Cartan lemma \cite{Clelland2017}, we can write $\omega_{13}$ and $\omega_{23}$ as
\begin{equation}\label{2.12}
\omega_{13}=a\omega_1+b\omega_2, \quad \omega_{23}=b\omega_1+c\omega_2,
\end{equation}
with $a$, $b$, $c$ differentiable functions, whose geometric interpretation is as follows: $a$ and $c$ are the normal curvatures of $S$ in the directions of $\textbf{e}_1$ and $\textbf{e}_2$, respectively; $b$ (resp., $-b$) is the geodesic torsion in the direction of $\textbf{e}_1$ (resp., $\textbf{e}_2$).
Therefore the equation (\ref{2.10}) reduces to
\begin{equation}\label{2.13}
d\omega_{12}=\omega_{1}\wedge\omega_{2},
\end{equation}
with Gauss equation
\begin{equation}\label{2.14}
ac-b^2=-1
\end{equation}
being the Gaussian curvature of $S$ in terms of its extrinsic geometry.

According to the theorem mentioned in \cite{Kahouadji 2016}, the Codazzi equation (\ref{2.11}) can be rewritten using the components $f_{ij}$ of the 1-forms $\omega_1, \omega_2,\omega_3:=\omega_{12}$ in the following form
 \begin{equation}\label{2.15}
 f_{11}D_t a+f_{21}D_t b-f_{12}D_x a-f_{22}D_x b-2b\Delta_{13}+(a-c)\Delta_{23}=0,
 \end{equation}
\begin{equation}\label{2.16}
f_{11}D_t b+f_{21}D_t c-f_{12}D_x b-f_{22}D_x c+(a-c)\Delta_{13}+2b\Delta_{23}=0,
\end{equation}
where
\begin{equation}\label{2.17}
\Delta_{ij}=f_{i1}f_{j2}-f_{j1}f_{i2}
\end{equation}
with
\begin{equation}\label{2.18}
\Delta_{13}^2+\Delta_{23}^2\neq 0.
\end{equation}
Moreover, in view of (\ref{1.8}) and (\ref{2.12}), the second fundamental form of local isometric immersions of surfaces determined by the solutions of an equation describing pseudospherical surfaces has the form
\begin{equation}\label{2.19}
II=a_{1}dx^2+2a_{2}dxdt+a_{3}dt^2,
\end{equation}
with
\begin{equation}\label{2.20}
\left\{\begin{aligned}
      a_{1}& = af_{11}^2+2bf_{11}f_{21}+cf_{21}^2,\\
      a_{2}&=af_{11}f_{12}+b(f_{11}f_{22}+f_{21}f_{12})+cf_{21}f_{22},\\
      a_{3}&=af_{12}^2+2bf_{12}f_{22}+cf_{22}^2.
       \end{aligned}
\right.
\end{equation}

In view of Bonnet theorem, the local isometric immersion of the pseudospherical surfaces characterized by the space of solutions of an equation describing pseudospherical surfaces exists if and only if there exists solutions $a$, $b$, $c$ of (\ref{2.14})-(\ref{2.16}), which are called the coefficients of the second fundamental form. In this paper, we will discuss the existence and dependence of these coefficients for the case of equations given by Theorem 2.2-2.5, under the assumption that they depend on $x$, $t$, $u_0$ and finite oder derivatives of $u_0$ with respect to $x$ and $t$.

\subsection{The classification of third order differential equations describing pseudospherical surfaces}

\indent \indent In our earlier work, we established a complete classification of equations (\ref{1.7}) from the flatness of connection 1-forms, which describe pseudospherical surfaces. We obtained five classes of equations determined by some differential functions. Here, we state these classification results without proof.

\begin{Lemma}\label{lemma2.1}
A partial differential equation of the form
\begin{equation}\label{2.21}
	u_{0,t} - u_{2,t} = \lambda u_{0}^2 u_{3} + G(u_{0}, u_{1}, u_{2}), \quad \lambda\in\mathbb{R}, \quad G \neq 0,
\end{equation}
with associated 1-forms $\omega_i=f_{i1}dx+f_{i2}dt, 1\leq i\leq3$, where the coefficient functions $f_{ij}=f_{ij}(u_0,u_1,\dots,u_k)$ satisfying the condition (\ref{1.9}), describes pseudospherical surfaces if and only if $f_{ij}$ and $G$ satisfy the following conditions:
\begin{equation}\label{2.22}
f_{i1,u_1}=0,\quad f_{i1,u_k}=f_{i2,u_k}=0,\quad 3\leq k \leq m,\quad m\in\mathbb{Z},
\end{equation}
\begin{equation}\label{2.23}
f_{i1,u_0}+f_{i1,u_2}=0,
\end{equation}
\begin{equation}\label{2.24}
f_{i2}=-\lambda u_{0}^{2}f_{i1}+\phi_{i},
\end{equation}
where $\phi_{i}=\phi_{i}(u_{0},u_{1})$ are real differential functions of $u_{0},u_{1}$ satisfying
\begin{equation}\label{2.25}
-Gf_{11,u_0}+(-2\lambda u_0 f_{11}-\lambda u_0^2 f_{11,u_0}+\phi_{1,u_0})u_{1}+\phi_{1,u_1}u_{2}+Mf_{11}+N=0,
\end{equation}
\begin{equation}\label{2.26}
Qf_{11}+L_{2,u_0}u_{1}+L_{2,u_1}u_{2}-2\lambda \eta_{2}u_0 u_{1}-\mu_{2}N+\eta_{3}\phi_{1}=0,
\end{equation}
\begin{equation}\label{2.27}
-(\delta L_2+\mu_{3}M)f_{11}+L_{3,u_0}u_{1}+L_{3,u_1}u_{2}-2\lambda \eta_{3}u_0 u_{1}-\mu_{3}N+\delta\eta_{2}\phi_{1}=0,
\end{equation}
\begin{equation}\label{2.28}
-L_2 f_{11}+\eta_{2}\phi_{1}\neq 0,
\end{equation}
with
\begin{equation}\label{2.29}
\begin{split}
 & L_p= L_p(u_0, u_1) := \phi_{p} - \mu_p\phi_{1},\quad p=2,3, \\
 & M=M(u_0,u_1):=\mu_{2}\phi_{3}-\mu_{3}\phi_{2}, \\
 & N=N(u_0,u_1):=\eta_{2}\phi_{3}-\eta_{3}\phi_{2}, \\
 & Q = Q(u_0, u_1) := -(L_3+\mu_2 M), \\
 & \gamma := \mu_2\mu_3\eta_2 - (1 + \mu_2^2)\eta_3.
\end{split}
\end{equation}
\end{Lemma}

\begin{Theorem}\label{theorem2.2}
Consider a partial differential equation of the form (\ref{2.21}) which describes pseudospherical surfaces, with coefficient functions $f_{ij}=f_{ij}(u_0,u_1,\dots, u_k)$ satisfying (\ref{1.9}) and (\ref{2.22})-(\ref{2.29}) with $Q =L_2= 0,\gamma=0$, if and only if the equation (\ref{2.21}) can be written in the form
\begin{equation}\label{2.30}
u_{0,t}-u_{2,t}=\frac{1}{f'}\left(\phi_{1,u_0}u_{1}+\phi_{1,u_1}u_{2}\pm \frac{\eta_2}{\sqrt{1+\mu_{2}^2}}\phi_{1}\right),\quad \eta_2\neq0.
\end{equation}
Moreover, associated 1-forms are
\begin{equation}\label{2.31}
\begin{aligned}
&\omega_{1}=f\,dx+\phi_{1}\,dt,
\\&\omega_{2}=(\mu_{2}f_{11}+\eta_{2})\,dx+\mu_{2}\phi_{1}\,dt,
\\&\omega_{3}=\pm \left(\sqrt{1+\mu_{2}^2}f_{11}+\frac{\mu_{2} \eta_2}{\sqrt{1+\mu_{2}^2}}\right)\,dx\pm \sqrt{1+\mu_{2}^2}\phi_{1}\,dt,
\end{aligned}
\end{equation}
where $\phi_{1} \neq 0$ and $f = f(u_0 - u_2)$ is a differentiable function such that $f' \neq 0$.
\end{Theorem}

\begin{Theorem}\label{theorem2.3}
Consider a partial differential equation of the form (\ref{2.21}) which describes pseudospherical surfaces, with coefficient functions $f_{ij}=f_{ij}(u_0,u_1,\dots, u_k)$ satisfying (\ref{1.9}) and (\ref{2.22})-(\ref{2.29}) with $Q =L_2=0,\gamma\neq 0$, if and only if the equation (\ref{2.21}) can be written in the form
\begin{equation}\label{2.32}
u_{0,t}-u_{2,t}= \lambda u_0^2 u_{3}
-\frac{\lambda}{f'}\left[2u_0 u_{1}f+u_0^{2}u_{1}f'+\frac{2\eta_{2}}{\gamma}(u_{1}^{2}+u_0 u_{2}+(\mu_{3}\eta_{2}-\mu_{2}\eta_{3})u_0 u_{1})\right],\quad \lambda \eta_2\neq0.
\end{equation}
Moreover, associated 1-forms are
\begin{equation}\label{2.33}
\begin{aligned}
&\omega_{1}=f\,dx-\lambda\left( u_0^{2}f+\frac{2}{\gamma}\eta_{2} u_0 u_{1}\right)\,dt,
\\&\omega_{2}=(\mu_{2}f+\eta_{2})\,dx-\lambda \left(u_0^{2}f_{21}+\frac{2}{\gamma}\mu_{2}\eta_{2} u_0 u_{1}\right)\,dt,
\\&\omega_{3}=(\mu_{3}f+\eta_{3})\,dx-\lambda \left(u_0^{2}f_{31}+\frac{2}{\gamma}\mu_{3}\eta_{2} u_0 u_{1}\right)\,dt,
\end{aligned}
\end{equation}
where $\eta_{2}^{2}-\eta_{3}^{2}-(\mu_{2}\eta_{3}-\mu_{3}\eta_{2})^{2}=0$ and $f=f(u_0- u_2)$ is a differentiable function with $f'\neq0$.
\end{Theorem}

\begin{Theorem}\label{theorem2.4}
Consider a partial differential equation of the form (\ref{2.21}) which describes pseudospherical surfaces, with coefficient functions $f_{ij}=f_{ij}(u_0,u_1,\dots, u_k)$ satisfying (\ref{1.9}) and (\ref{2.22})-(\ref{2.29}) with $Q =0, L_2\neq 0,\gamma=0$, if and only if the equation (\ref{2.21}) can be written in the form
\begin{equation}\label{2.34}
\begin{split}
u_{0,t}-u_{2,t}= &\lambda u_0^2 u_{3}+\frac{1}{f'}\left[u_{1}\phi_{1,u_0}+u_{2}\phi_{1,u_1}-\lambda u_0^{2}u_{1}f'\pm\frac{\eta_2}{\sqrt{1+\mu_{2}^2}}\phi_{1}\right.
\\&\left.-\left(2\lambda u_0 u_{1}\pm\frac { \eta_2 }{\sqrt{1+\mu_{2}^2}}\lambda u_0^{2}\pm\frac{C_1}{\sqrt{1+\mu_{2}^2}}\right)f\right], \quad C_1
\in \mathbb{R}.
\end{split}
\end{equation}
Moreover, associated 1-forms are
\begin{equation}\label{2.35}
\begin{aligned}
&\omega_{1}=f\,dx-(\lambda u_0^{2}f-\phi_{1})\,dt,
\\&\omega_{2}=(\mu_{2}f+\eta_{2})\,dx-(\lambda \mu_{2}u_0^{2}f-\mu_{2} \phi_{1}-C_1)\,dt,
\\&\omega_{3}=\pm \left(\sqrt{1+\mu_{2}^2}f+ \frac{\mu_{2} \eta_2}{\sqrt{1+\mu_{2}^2}}\right)\,dx\mp \sqrt{1+\mu_{2}^2}\left(\lambda u_0^{2}f- \phi_{1}- \frac{\mu_{2}C_1}{1+\mu_{2}^2}\right)\,dt,
\end{aligned}
\end{equation}
where $(\lambda \eta_2)^{2}+C_1^{2} \neq 0$ and $f=f(u_0-u_2)$ is a differentiable function such that $f'\neq 0$.
\end{Theorem}

\begin{Theorem}\label{theorem2.5}
Consider a partial differential equation of the form (\ref{2.21}) which describes pseudospherical surfaces, with coefficient functions $f_{ij}=f_{ij}(u_0,u_1,\dots, u_k)$ satisfying (\ref{1.9}) and (\ref{2.22})-(\ref{2.29}) with $Q\neq0 ,L_2\neq 0,\gamma\neq0$, if and only if the equation (\ref{2.21}) can be written in one of the following two forms
\\(i)
\begin{equation}\label{2.36}
\begin{split}
u_{0,t}-u_{2,t}=& \lambda u_0^2 u_{3}+\lambda \left[-5u_0^{2}u_{1}+4u_0 u_{1}u_{2}+\left(2\zeta_{1}-\frac{4}{\theta}\right)u_0 u_{1}
-\frac{2}{\theta}u_{1}u_{2}+\frac{2\zeta_{1}}{\theta}u_{1}\right]
\\&+\left(\theta u_{1}^{3}+2u_0 u_{1}+u_{1}u_{2}-\zeta_{1}u_{1}\right)\theta C_2 e^{\theta u_0},\quad 0\neq\nu\theta,\sigma,C_2\in\mathbb{R},\quad \lambda^{2}+C_2^{2}\neq 0,
\end{split}
\end{equation}
and associated 1-forms are
\begin{equation}\label{2.37}
\begin{aligned}
\omega_{1}=&[\nu(u_0-u_2)-\sigma]\,dx-\left[\lambda u_0^{2}f_{11}+\frac{\nu}{\theta}\left(2\lambda -\theta^{2}C_2 e^{\theta u_0}\right)u_{1}^{2}\right.\\
&+\left.\left(\frac{2\lambda}{\theta}-\theta C_2 e^{\theta u_0}+2\lambda u_0\right)
\left(\frac{\nu u_0-\sigma}{\theta}\pm\left(\mu_{2}-\frac{\nu\eta_{2}}{\theta}\right)\frac{u_1}{\sqrt{1+\mu_{2}^2}}\right)\right]\,dt,\\
\omega_{2}=&(\mu_{2}f_{11}+\eta_{2})\,dx+\left[\mu_{2}f_{12}-\lambda \eta_{2}u_0^{2}
+\left(\frac{2\lambda}{\theta}-\theta C_2 e^{\theta u_0}+2\lambda u_0\right)\left(\pm \sqrt{1+\mu_{2}^2}u_{1}-\frac{\eta_{2}}{\theta}\right)\right]\,dt,\\
\omega_{3}=&\pm \left(\sqrt{1+\mu_{2}^2}f_{11}\pm\eta_3\right)\,dx
\\&\pm\left[ \sqrt{1+\mu_{2}^2}f_{12}\mp\lambda \eta_{3}u_0^{2}\pm\left(\frac{2\lambda}{\theta}-\theta C_2 e^{\theta u_0}+2\lambda u_0 \right)
\left(\mu_{2}u_{1}-\eta_3\right)\right]\,dt,
\end{aligned}
\end{equation}
with $$\zeta_{1}=\frac{2\sigma}{\nu}-\frac{1}{\theta}-\frac{\theta}{\nu^{2}(1+\mu_{2}^{2})}-\frac{\eta_{2}(2\theta \mu_{2}+\nu\eta_{2})}{\theta \nu(1+\mu_{2}^{2})},$$
\\or
\\(ii)
\begin{equation}\label{2.38}
\begin{split}
u_{0,t}-u_{2,t}=& \lambda u_0^2 u_{3}+\lambda \left[-3u_0^{2}u_{1}+2u_0 u_{1}u_{2}+2\zeta_{2}u_0 u_{1}\mp \frac{2}{\tau}(u_{1}^{2}+u_0 u_{2})\right]
\\&+\tau \left(\tau u_0 u_{2}\pm u_{1}-\zeta_{2}\tau u_{2}\right)\varphi e^{\pm \tau u_1}+\varphi^{''}u_{1}^{2}e^{\pm \tau u_1}
\\&\pm(\tau u_0 u_{1}+\tau u_{1}u_{2}\pm u_{2}-\zeta_{2}\tau u_{1})\varphi^{'}e^{\pm \tau u_1}
,\quad 0<\tau,\nu,\sigma\in \mathbb{R},\quad\nu\eta_2\neq0,
\end{split}
\end{equation}
and associated 1-forms are
\begin{equation}\label{2.39}
\begin{split}
\omega_{1}=&(\nu(u_0-u_2)-\sigma)\,dx
-\left[\lambda u_0^{2}f_{11}-(\pm\tau(\nu u_0-\sigma)\varphi+\nu\varphi 'u_{1})e^{\pm \tau u_1}\pm \frac{2\lambda \nu}{\tau}u_0 u_{1}\right]\,dt,\\
\omega_{2}=&(\mu_{2}f_{11}+\eta_{2})\,dx+
\left(\mu_{2}f_{12}-\lambda \eta_{2}u_0^{2}\pm \tau \eta_{2}\varphi e^{\pm \tau u_1}\right)\,dt,\\
\omega_{3}=&\pm\tau \left[\left(\frac{\sigma}{\nu}-\zeta_{2}\right)\left(\frac{1+\mu_{2}^{2}}{\eta_2}f_{11}+\mu_{2}\right)- \frac{1}{\nu}f_{21}\right]\,dx\\
&\pm\tau\left[ \left(\frac{\sigma}{\nu}-\zeta_{2}\right)\left[\frac{1+\mu_{2}^{2}}{\eta_2}f_{12}-\mu_{2}\left(\lambda u_0^{2}\mp\tau \varphi e^{\pm\tau u_1}\right)\right]- \frac{1}{\nu}f_{22}\right]\,dt,
\end{split}
\end{equation}
where $\varphi(u_0) \neq 0$ is arbitrary differentiable function and $$\zeta_{2}=\frac{\sigma}{\nu}\mp\frac{\mu_{3}\eta_{2}-\mu_{2}\eta_{3}}{\tau}.$$	
\end{Theorem}

\section{Proof of Theorem 1.1}

\indent \indent The present section will concentrate on the proof of Theorem 1.1, which consists of several technical lemmas and propositions. Our analysis will specifically address the existence of solutions to the system of equations (\ref{2.14})-(\ref{2.16}) that depend on a jet of finite order of $u_0$, for each classes of equations in Theorem 2.2-2.5. It's worth noting that the coefficients $f_{ij}$, as presented in Theorems 2.2-2.5, only depend on $u_0$, $u_1$, and $u_2$.

\begin{Lemma}\label{lemma3.1}
Consider a partial differential equation of the form (\ref{2.21}) which describes pseudospherical surfaces, given by Theorems 2.2-2.5. Assume there exists a local isometric immersion of the pseudospherical surface, determined by a solution $u(x,t)$ of the equation (\ref{2.21}), for which the coefficients $a$, $b$, $c$ of the second fundamental form depend on $x$, $t$, $u_0, \dots, u_l$, $w_1, \dots, w_m$, $v_1, \dots, v_n$, $1 \leq l,m,n < \infty$. Then $ac \neq 0$ on any open set.
\end{Lemma}
The proof of this lemma follows analogous arguments to those presented in \cite{Silva2016} (Lemma 3.1), and thus is omitted here for conciseness.

\hspace*{\fill}\\

The pseudospherical surfaces determined by the space of solutions of the equation (\ref{2.21}) given by Theorems 2.2-2.5, admit a local isometric immersion with the coefficients $a$, $b$, $c$ depend on a jet of finite order of $u_0$,  if and only if the system of equations (\ref{2.14})-(\ref{2.16}) are satisfied. By substituting the total derivatives with respect to $x$ and $t$ provided by (\ref{2.2}) and (\ref{2.3}), we rewrite the equations (\ref{2.15}) and (\ref{2.16}) as
\begin{equation}\label{3.1}
\begin{aligned}
&f_{11}a_t+f_{21}b_t-f_{12}a_x-f_{22}b_x-2b(f_{11}f_{32}-f_{31}f_{12})+(a-c)(f_{21}f_{32}-f_{31}f_{22})\\
&+\sum_{i=2}^l (f_{11}a_{u_i}+f_{21}b_{u_i})\partial_x^{i-2}(u_{0,t}-F)-\sum_{i=0}^l (f_{12}a_{u_i}+f_{22}b_{u_i}) u_{i+1}\\
&+\sum_{j=0}^m (f_{11}a_{w_j}+f_{21}b_{w_j})w_{j+1}-\sum_{j=1}^m (f_{12}a_{w_j}+f_{22}b_{w_j}) w_{j,x}\\
&+\sum_{k=0}^n (f_{11}a_{v_k}+f_{21}b_{v_k})v_{k+1}-\sum_{k=1}^n (f_{12}a_{v_k}+f_{22}b_{v_k}) v_{k,x}=0,
\end{aligned}
\end{equation}
and
\begin{equation}\label{3.2}
\begin{aligned}
&f_{11}b_t+f_{21}c_t-f_{12}b_x-f_{22}c_x+(a-c)(f_{11}f_{32}-f_{31}f_{12})+2b(f_{21}f_{32}-f_{31}f_{22})\\
&+\sum_{i=2}^l (f_{11}b_{u_i}+f_{21}c_{u_i})\partial_x^{i-2}(u_{0,t}-F)-\sum_{i=0}^l (f_{12}b_{u_i}+f_{22}c_{u_i}) u_{i+1}\\
&+\sum_{j=0}^m (f_{11}b_{w_j}+f_{21}c_{w_j})w_{j+1}-\sum_{j=1}^m (f_{12}b_{w_j}+f_{22}c_{w_j}) w_{j,x}\\
&+\sum_{k=0}^n (f_{11}b_{v_k}+f_{21}c_{v_k})v_{k+1}-\sum_{k=1}^n (f_{12}b_{v_k}+f_{22}c_{v_k}) v_{k,x}=0.
\end{aligned}
\end{equation}

Without loss of generality, we may assume $m=n$ throughout our analysis, as the cases where $m<n$ or $m>n$ can be reduced to the case $m=n$. In fact, if $m<n$, i.e. $n\geq m+1$. Differentiating (\ref{3.1}), (\ref{3.2}) and (\ref{2.14}) with respect to $v_{n+1}$ implies $a_{v_n} = b_{v_n} = c_{v_n} = 0$. Successive differentiation with respect to $v_n, v_{n-1},\dots,v_{(m+1)+1}$ implies $a_{v_{n-1}} =\dots = a_{v_{m+1}} = 0$, $b_{v_{n-1}} = \dots = b_{v_{m+1}} = 0$ and $c_{v_{n-1}} = \dots = c_{v_{m+1}} = 0$. On the other hand, if $m>n$, i.e. $m\geq n+1$. Differentiating (\ref{3.1}), (\ref{3.2}) and (\ref{2.14}) with respect to $w_{m+1}$ implies $a_{w_m} = b_{w_m} =c_{w_m} = 0$. Successive differentiation with respect to $w_m, w_{m-1},\dots,w_{(n+1)+1}$ implies $a_{w_{m-1}} =\dots = a_{w_{n+1}} = 0$, $b_{w_{m-1}} = \dots = b_{w_{n+1}} = 0$ and $c_{w_{m-1}} = \dots = c_{w_{n+1}} = 0$. Hence, $a$, $b$ and $c$ are functions of $x$, $t$, $u_0$, $u_1$, $\dots$, $u_l$, $w_1$, $\dots$, $w_n$, $v_1$, $\dots$, $v_n$.

Back to the equations (\ref{3.1}) and (\ref{3.2}), differentiating both equations with respect to $v_{n+1}$ and $w_{n+1}$, we find
\begin{equation}\label{3.3}
\begin{aligned}
&f_{11}a_{v_n}+f_{21}b_{v_n}=0,\quad &f_{11}a_{w_n}+f_{21}b_{w_n}=0,\\
&f_{11}b_{v_n}+f_{21}c_{v_n}=0,\quad &f_{11}b_{w_n}+f_{21}c_{w_n}=0.
\end{aligned}
\end{equation}
The following lemmas will be discussed in two separate cases: $f_{21}=0$ and $f_{21}\neq 0$. Moreover, if $f_{21}\neq 0$ on a non-empty open set, then
\begin{equation}\label{3.4}
\begin{aligned}
&b_{v_n}=-\frac{f_{11}}{f_{21}}a_{v_n},\quad &b_{w_n}&=-\frac{f_{11}}{f_{21}}a_{w_n},\\
&c_{v_n}=\left(\frac{f_{11}}{f_{21}}\right)^2 a_{v_n}, \quad &c_{w_n}&=\left(\frac{f_{11}}{f_{21}}\right)^2 a_{w_n}.
\end{aligned}
\end{equation}
The derivative of Gauss equation (\ref{2.14}) with respect to $v_n$ and $w_n$ returns
\begin{equation}\label{3.5}
\left[c+\left(\frac{f_{11}}{f_{21}}\right)^2 a+2\frac{f_{11}}{f_{21}}b\right]a_{v_n}=0,\quad \left[c+\left(\frac{f_{11}}{f_{21}}\right)^2 a+2\frac{f_{11}}{f_{21}}b\right]a_{w_n}=0,
\end{equation}
and thus, the lemmas will be proceeded by further distinguishing two subcases:
$$ c+\left(\frac{f_{11}}{f_{21}}\right)^2 a+2b\frac{f_{11}}{f_{21}}\neq 0 \quad and \quad c+\left(\frac{f_{11}}{f_{21}}\right)^2 a+2b\frac{f_{11}}{f_{21}}=0.$$

\begin{Lemma}\label{lemma3.2}
Consider a partial differential equation of the form (\ref{2.21}) which describes pseudospherical surfaces, given by Theorems 2.2-2.5. Assume there exists a local isometric immersion of the pseudospherical surface, determined by a solution $u(x,t)$ of the equation (\ref{2.21}), for which the coefficients $a$, $b$, $c$ of the second fundamental form depend on $x$, $t$, $u_0, \dots, u_l$, $w_1, \dots, w_n$, $v_1,\dots, v_n$, $1 \leq l,n < \infty$. If $f_{21}=0$ on a non-empty open set, then $a$, $b$, and $c$ are universal functions of $x$ and $t$, independent of $u_0$.
\end{Lemma}

\begin{Proof}
The case $f_{21} = 0$ is only achieved in (\ref{2.35}) and (\ref{2.37}) under the parameter constraints $\mu_2 = \eta_2 = 0$. Crucially, for both classes, $\phi_{2}$ is everywhere non-vanishing. If $f_{21} = 0$, the equations (\ref{3.1}) and (\ref{3.2}) are simplified to
\begin{equation}\label{3.6}
\begin{aligned}
&f_{11}a_t-f_{12}a_x-f_{22}b_x-2b(f_{11}f_{32}-f_{31}f_{12})-(a-c)f_{31}f_{22}\\
&+\sum_{i=2}^l f_{11}a_{u_i}\partial_x^{i-2}(u_{0,t}-F)-\sum_{i=0}^l (f_{12}a_{u_i}+f_{22}b_{u_i}) u_{i+1}
+\sum_{j=0}^n f_{11}a_{w_j}w_{j+1}\\
&-\sum_{j=1}^n (f_{12}a_{w_j}+f_{22}b_{w_j}) w_{j,x}
+\sum_{k=0}^n f_{11}a_{v_k}v_{k+1}-\sum_{k=1}^n (f_{12}a_{v_k}+f_{22}b_{v_k}) v_{k,x}=0,
\end{aligned}
\end{equation}
and
\begin{equation}\label{3.7}
\begin{aligned}
&f_{11}b_t-f_{12}b_x-f_{22}c_x+(a-c)(f_{11}f_{32}-f_{31}f_{12})-2bf_{31}f_{22}\\
&+\sum_{i=2}^l f_{11}b_{u_i}\partial_x^{i-2}(u_{0,t}-F)-\sum_{i=0}^l (f_{12}b_{u_i}+f_{22}c_{u_i}) u_{i+1}
+\sum_{j=0}^n f_{11}b_{w_j}w_{j+1}\\
&-\sum_{j=1}^n (f_{12}b_{w_j}+f_{22}c_{w_j}) w_{j,x}
+\sum_{k=0}^n f_{11}b_{v_k}v_{k+1}-\sum_{k=1}^n (f_{12}b_{v_k}+f_{22}c_{v_k}) v_{k,x}=0.
\end{aligned}
\end{equation}

Suppose $l=1$. Successive differentiation of (\ref{3.6}) and (\ref{3.7}) with respect to $v_{n+1},\dots,v_1$ and $w_{n+1},\dots,w_1$ and of the Gauss equation (\ref{2.14}) with respect to $v_n,\dots,v_0$, combined with the condition $f_{11}\neq 0$, implies $a_{v_k} = b_{v_k} = c_{v_k} = 0$ and $a_{w_j} = b_{w_j} = c_{w_j} = 0$ for $j,k = 0, 1,\dots,n$. Hence, $a$, $b$ and $c$ are universal functions of $x$ and $t$, independent of $u_0$.

Suppose $l\geq 2$. Taking successive differentiation of (\ref{3.6}) and (\ref{3.7}) with respect to $v_{n+1},\dots,v_2$ and $w_{n+1},\dots,w_2$ and of the Gauss equation (\ref{2.14}) with respect to $v_n,\dots,v_1$, combined with the condition $f_{11}\neq 0$, implies $a_{v_k} = b_{v_k} = c_{v_k} = 0$ and $a_{w_j} = b_{w_j} = c_{w_j} = 0$ for $j,k = 1, 2,\dots,n$. Thus, $a$, $b$ and $c$ depend on $x$, $t$, $u_0$, $u_1$, $\dots$, $u_l$. Moreover, (\ref{3.6}) and (\ref{3.7}) become to
\begin{equation}\label{3.8}
\begin{aligned}
&f_{11}a_t-f_{12}a_x-f_{22}b_x-2b(f_{11}f_{32}-f_{31}f_{12})-(a-c)f_{31}f_{22}\\
&+\sum_{i=2}^l f_{11}a_{u_i}\partial_x^{i-2}(u_{0,t}-F)-\sum_{i=0}^l (f_{12}a_{u_i}+f_{22}b_{u_i}) u_{i+1}+f_{11}a_{w_0}w_1+f_{11}a_{v_0}v_1=0,
\end{aligned}
\end{equation}
and
\begin{equation}\label{3.9}
\begin{aligned}
&f_{11}b_t-f_{12}b_x-f_{22}c_x+(a-c)(f_{11}f_{32}-f_{31}f_{12})-2bf_{31}f_{22}\\
&+\sum_{i=2}^l f_{11}b_{u_i}\partial_x^{i-2}(u_{0,t}-F)-\sum_{i=0}^l (f_{12}b_{u_i}+f_{22}c_{u_i}) u_{i+1}+ f_{11}b_{w_0}w_1+f_{11}b_{v_0}v_1=0.
\end{aligned}
\end{equation}
Performing differentiation with respect to $u_{l+1}$ on (\ref{3.8}) and (\ref{3.9}), we derive
\begin{equation}\label{3.10}
\phi_{1}a_{u_l}+\phi_{2}b_{u_l}=0, \quad \phi_{1}b_{u_l}+\phi_{2}c_{u_l}=0.
\end{equation}
The $u_l$ derivative of the Gauss equation (\ref{2.14}) yields $a_{u_l}c+ac_{u_l}-2bb_{u_l}=0$. Taking into account the equations (\ref{3.10}) in the latter, we obtain
\begin{equation}\label{3.11}
\left[c+\left(\frac{\phi_{1}}{\phi_{2}}\right)^2a+2b\frac{\phi_{1}}{\phi_{2}}\right]a_{u_l}=0.
\end{equation}

If the expression within brackets in (\ref{3.11}) is non-vanishing, then from (\ref{3.11}), we get $a_{u_l}=0$. Consequently, (\ref{3.10}) implies that $b_{u_l}=c_{u_l}=0$. By successively differentiating equations (\ref{3.8}) and (\ref{3.9}) with respect to $u_l,\dots,u_3$, we have $a_{u_i}=b_{u_i}=c_{u_i}=0$ for $i=2,\dots, l$. Furthermore, differentiation of equations (\ref{3.6}) and (\ref{3.7}) with respect to $v_1$ and $w_1$ leads to $a_{v_0}=b_{v_0}=0$ and $a_{w_0}=b_{w_0}=0$. Finally, differentiating the Gauss equation (\ref{2.14}) with respect to $w_0$ and $v_0$, and utilizing the condition $a\neq0$, we conclude $c_{v_0}=c_{w_0}=0$. Hence, $a$, $b$ and $c$ are universal functions of $x$ and $t$, independent of $u_0$.

If the expression within brackets in (\ref{3.11}) is vanishing, it follows from (\ref{2.14}) that
\begin{equation}\label{3.12}
  b=\pm 1-\frac{\phi_{1}}{\phi_{2}}a,\quad c=\left(\frac{\phi_{1}}{\phi_{2}}\right)^2 a\mp2 \frac{\phi_{1}}{\phi_{2}}.
\end{equation}
Moreover, the following identities hold
\begin{equation}\label{3.13}
\begin{aligned}
&D_t b=-\frac{\phi_{1}}{\phi_{2}}D_t a-aD_t \left(\frac{\phi_{1}}{\phi_{2}}\right), &D_t c&=\left(\frac{\phi_{1}}{\phi_{2}}\right)^2 D_t a+2 \left(\frac{\phi_{1}}{\phi_{2}}a\mp1\right)D_t \left(\frac{\phi_{1}}{\phi_{2}}\right),\\
&D_x b=-\frac{\phi_{1}}{\phi_{2}}D_x a-aD_x \left(\frac{\phi_{1}}{\phi_{2}}\right), &D_x c&=\left(\frac{\phi_{1}}{\phi_{2}}\right)^2 D_x a+2 \left(\frac{\phi_{1}}{\phi_{2}}a\mp1\right)D_x \left(\frac{\phi_{1}}{\phi_{2}}\right),
\end{aligned}
\end{equation}
Then the equations (\ref{2.15}) and (\ref{2.16}) can be rewritten as
\begin{equation}\label{3.14}
f_{11}(D_t a+\lambda u_0^2D_x a)+af_{22} D_x \left(\frac{\phi_{1}}{\phi_{2}}\right)-2b\Delta_{13}+(a-c)\Delta_{23}=0,
\end{equation}
and
\begin{equation}\label{3.15}
\begin{split}
&-\frac{\phi_{1}}{\phi_{2}}f_{11}(D_t a+\lambda u_0^2D_x a)
-\left[a\left(\lambda u_0^2 f_{11}+f_{22}\frac{\phi_{1}}{\phi_{2}}\right)\mp2 f_{22}\right]D_x \left(\frac{\phi_{1}}{\phi_{2}}\right)\\&-f_{11}a D_t \left(\frac{\phi_{1}}{\phi_{2}}\right)+(a-c)\Delta_{13}+2b\Delta_{23}=0.
\end{split}
\end{equation}
Adding (\ref{3.14}) multiplied by $\phi_{1}/\phi_{2}$ with (\ref{3.15}), we have
\begin{equation}\label{3.16}
\begin{split}
&-f_{11}a \left[\left(\frac{\phi_{1}}{\phi_{2}}\right)_{u_0}u_{0,t}+\left(\frac{\phi_{1}}{\phi_{2}}\right)_{u_1} u_{1,t}\right]-(\lambda u_0^2 af_{11}\mp 2f_{22})D_x \left(\frac{\phi_{1}}{\phi_{2}}\right)\\
&+\left(a-c-2\frac{\phi_{1}}{\phi_{2}}b\right)\Delta_{13}+\left[\frac{\phi_{1}}{\phi_{2}}(a-c)+2b\right]\Delta_{23}=0,
\end{split}
\end{equation}
and then differentiating (\ref{3.16}) with respect to $v_1=u_{1,t}$ and $w_1=u_{0,t}$, we find
\begin{equation}\label{3.17}
f_{11}a \left(\frac{\phi_{1}}{\phi_{2}}\right)_{u_1}=0, \quad f_{11}a \left(\frac{\phi_{1}}{\phi_{2}}\right)_{u_0}=0,
\end{equation}
which imply that $\phi_{2}/\phi_{1}=\mu$, $\mu \in \mathbb{R} \setminus \{0\}$. Note that $L_2=\phi_{2}-\mu\phi_{1}=0$ can not happen in (\ref{2.35}) or (\ref{2.37}), which yields a contradiction.

Therefore, when $f_{21}=0$ on a non-empty open set, $a$, $b$ and $c$ are universal functions of $x$ and $t$, independent of $u_0$.
$\hfill\square$
\end{Proof}

\begin{Lemma}\label{lemma3.3}
Consider a partial differential equation of the form (\ref{2.21}) which describes pseudospherical surfaces, given by Theorems 2.2-2.5. Assume there exists a local isometric immersion of the pseudospherical surface, determined by a solution $u(x,t)$ of the equation (\ref{2.21}), for which the coefficients $a$, $b$, $c$ of the second fundamental form depend on $x$, $t$, $u_0, \dots, u_l$, $w_1, \dots, w_n$, $v_1,\dots, v_n$, $1 \leq l,n < \infty$. If $f_{21}\neq0$ and
\begin{equation}\label{3.18}
c+\left(\frac{f_{11}}{f_{21}}\right)^2 a+2\frac{f_{11}}{f_{21}}b\neq 0
\end{equation}
on a non-empty open set, then $a$, $b$, and $c$ are universal functions of $x$ and $t$, independent of $u_0$.
\end{Lemma}

\begin{Proof}
If (\ref{3.18}) holds, then it follows from (\ref{3.5}) that $a_{v_n}=a_{w_n}=0$ and thus equations (\ref{3.4}) imply that $b_{v_n}=b_{w_n}=0$ and $b_{v_n}=b_{w_n}=0$.

Suppose $l=1$, similar to the previous proof, successive differentiation of (\ref{3.1}), (\ref{3.2}) and the Gauss equation (\ref{2.14}) leads to $a_{v_k} = b_{v_k} = c_{v_k} = 0$ and $a_{w_j} = b_{w_j} = c_{w_j} = 0$ for $j,k = 0, 1,\dots,n-1$. Hence, $a$, $b$, and $c$ are universal functions of $x$ and $t$, independent of $u_0$.

Suppose $l\geq 2$, applying analogous differentiation procedures to (\ref{3.1}), (\ref{3.2}) and the Gauss equation (\ref{2.14}) yields $a_{v_k} = b_{v_k} = c_{v_k} = 0$ and $a_{w_j} = b_{w_j} = c_{w_j} = 0$ for $j,k = 1, 2,\dots,n$. Thus, $a$, $b$ and $c$ are function of $x$, $t$, $u_0$, $u_1$, $\dots$, $u_l$.
Differentiating (\ref{3.1}) and (\ref{3.2}) with respect to $u_{l+1}$, in views of (\ref{2.24}), gives
\begin{equation}\label{3.19}
\phi_{1}a_{u_l}+\phi_{2}b_{u_l}=0, \quad \phi_{1}b_{u_l}+\phi_{2}c_{u_l}=0,
\end{equation}

 When $\phi_{2}=0$ with $\Delta_{12}\neq 0$, we have $\phi_{1}\neq 0$. Then (\ref{3.19}) gives $a_{u_l}=b_{u_l}=0$, while differentiating the Gauss equation (\ref{2.14}) with respect to $u_l$, via Lemma 3.1, yields $c_{u_l}=0$. Successive differentiation of (\ref{3.1}) and (\ref{3.2}) with respect to $u_l,\dots,u_3$ and (\ref{2.14}) with respect to $u_{l-1},\dots,u_2$ shows $a_{u_i} = b_{u_i} = c_{u_i} = 0$ for $i= 2, 3,\dots,l-1$, reducing (\ref{3.1}) and (\ref{3.2}) to
\begin{equation}\label{3.20}
\begin{aligned}
&f_{11}a_t+f_{21}b_t-f_{12}a_x-f_{22}b_x-2b(f_{11}f_{32}-f_{31}f_{12})+(a-c)(f_{21}f_{32}-f_{31}f_{22})\\
&-\sum_{i=0}^1 (f_{12}a_{u_i}+f_{22}b_{u_i}) u_{i+1}+ (f_{11}a_{w_0}+f_{21}b_{w_0})w_1+ (f_{11}a_{v_0}+f_{21}b_{v_0})v_1=0,
\end{aligned}
\end{equation}
and
\begin{equation}\label{3.21}
\begin{aligned}
&f_{11}b_t+f_{21}c_t-f_{12}b_x-f_{22}c_x+(a-c)(f_{11}f_{32}-f_{31}f_{12})+2b(f_{21}f_{32}-f_{31}f_{22})\\
&-\sum_{i=0}^1 (f_{12}b_{u_i}+f_{22}c_{u_i}) u_{i+1}+ (f_{11}b_{w_0}+f_{21}c_{w_0})w_1+ (f_{11}b_{v_0}+f_{21}c_{v_0})v_1=0.
\end{aligned}
\end{equation}
Differentiating (\ref{3.20}) and (\ref{3.21}) with respect to $v_1$ and $w_1$ leads to
\begin{equation}\label{3.22}
\begin{split}
&f_{11}a_{v_0}+f_{21}b_{v_0}=0,\quad f_{11}a_{w_0}+f_{21}b_{w_0}=0,\\
&f_{11}b_{v_0}+f_{21}c_{v_0}=0,\quad f_{11}b_{w_0}+f_{21}c_{w_0}=0.
\end{split}
\end{equation}
The derivative of the Gauss equation (\ref{2.14}) with respect to $v_0$ and $w_0$ returns
\begin{equation}\label{3.23}
\left[c+\left(\frac{f_{11}}{f_{21}}\right)^2 a+2\frac{f_{11}}{f_{21}}b\right]a_{v_0}=0,\quad \left[c+\left(\frac{f_{11}}{f_{21}}\right)^2 a+2\frac{f_{11}}{f_{21}}b\right]a_{w_0}=0,
\end{equation}
and it follows from (\ref{3.18}) that $a_{v_0}=a_{w_0}=0$. Moreover, from (\ref{3.22}) we get $b_{v_0}=b_{w_0}=0$ and $c_{v_0}=c_{w_0}=0$. Hence, $a$, $b$, and $c$ are universal functions of $x$ and $t$, independent of $u_0$.

When $\phi_{2}\neq0$, the $u_l$ derivative of the Gauss equation (\ref{2.14}) combined with (\ref{3.19}) gives
\begin{equation}\label{3.24}
\left[c+\left(\frac{\phi_{1}}{\phi_{2}}\right)^2a+2b\frac{\phi_{1}}{\phi_{2}}\right]a_{u_l}=0.
\end{equation}

If the expression within brackets in (\ref{3.24}) is non-vanishing, then $a_{u_l}=0$ and (\ref{3.19}) implies that $b_{u_l}=c_{u_l}=0$. Through successive differentiation of (\ref{3.1}) and (\ref{3.2}) with respect to $u_l,\dots,u_3$, we establish $a_{u_i}=b_{u_i}=c_{u_i}=0$ for $i=2,\dots,l$. Furthermore, differentiation of (\ref{3.1}) and (\ref{3.2}) with respect to $v_1$ and $w_1$ leads to $a_{v_0}=b_{v_0}=0$ and $a_{w_0}=b_{w_0}=0$. Finally, differentiating the Gauss equation (\ref{2.14}) with respect to $w_0$ and $v_0$ and utilizing the condition $a\neq0$, we conclude $c_{v_0}=c_{w_0}=0$. Hence, $a$, $b$ and $c$ are universal functions of $x$ and $t$, independent of $u_0$.

If the expression within brackets in (\ref{3.24}) is vanishing, it follows from (\ref{2.14}) that
\begin{equation}\label{3.25}
  b=\pm 1-\frac{\phi_{1}}{\phi_{2}}a,\quad c=\left(\frac{\phi_{1}}{\phi_{2}}\right)^2 a\mp2 \frac{\phi_{1}}{\phi_{2}}.
\end{equation}
Then, by using (\ref{3.13}), the equations (\ref{2.15}) and (\ref{2.16}) become
\begin{equation}\label{3.26}
\frac{\Delta_{12}}{\phi_{2}}D_t a-af_{21}D_t \left(\frac{\phi_{1}}{\phi_{2}}\right)+\lambda u_0^2 \frac{\Delta_{12}}{\phi_{2}}D_x a+af_{22} D_x \left(\frac{\phi_{1}}{\phi_{2}}\right)-2b\Delta_{13}+(a-c)\Delta_{23}=0,
\end{equation}
and
\begin{equation}\label{3.27}
\begin{split}
&-\frac{\phi_{1}}{\phi_{2}}\frac{\Delta_{12}}{\phi_{2}}D_t a-a\frac{\Delta_{12}}{\phi_{2}}D_t \left(\frac{\phi_{1}}{\phi_{2}}\right)+f_{21}\left(a\frac{\phi_{1}}{\phi_{2}}\mp2\right)D_t \left(\frac{\phi_{1}}{\phi_{2}}\right)-\lambda u_0^2 \frac{\phi_{1}}{\phi_{2}}\frac{\Delta_{12}}{\phi_{2}}D_x a\\
&-\lambda u_0^2 a \frac{\Delta_{12}}{\phi_{2}}D_x \left(\frac{\phi_{1}}{\phi_{2}}\right)-f_{22}\left(a\frac{\phi_{1}}{\phi_{2}}\mp2\right)D_x \left(\frac{\phi_{1}}{\phi_{2}}\right)+(a-c)\Delta_{13}+2b\Delta_{23}=0.
\end{split}
\end{equation}
By adding the first equation multiplied by $\phi_{1}/\phi_{2}$ to the second equation, we have
\begin{equation}\label{3.28}
\begin{split}
&\left(-a\frac{\Delta_{12}}{\phi_{2}}\mp2f_{21}\right) \left[\left(\frac{\phi_{1}}{\phi_{2}}\right)_{u_0}u_{0,t}+\left(\frac{\phi_{1}}{\phi_{2}}\right)_{u_1} u_{1,t}\right]-(\lambda u_0^2 a\frac{\Delta_{12}}{\phi_{2}}\mp 2f_{22})D_x \left(\frac{\phi_{1}}{\phi_{2}}\right)\\
&+(a-c-2\frac{\phi_{1}}{\phi_{2}}b)\Delta_{13}+\left[\frac{\phi_{1}}{\phi_{2}}(a-c)+2b\right]\Delta_{23}=0,
\end{split}
\end{equation}
and then, by differentiating (\ref{3.28}) with respect to $v_1$ and $w_1$, we obtain
\begin{equation}\label{3.29}
\left(-a\frac{\Delta_{12}}{\phi_{2}}\mp2f_{21}\right)  \left(\frac{\phi_{1}}{\phi_{2}}\right)_{u_1}=0, \quad \left(-a\frac{\Delta_{12}}{\phi_{2}}\mp2f_{21}\right)  \left(\frac{\phi_{1}}{\phi_{2}}\right)_{u_0}=0.
\end{equation}

It is straightforward to verify that the bracketed expression in (\ref{3.29}) does not vanish. Consequently, we get $\phi_{2}/\phi_{1}=\mu$, $\mu \in \mathbb{R} \setminus \{0\}$, i.e. $L_2=\phi_{2}-\mu\phi_{1}=0$. This implies that the functions $f_{ij}$ satisfying are precisely those given by (\ref{2.31}) and (\ref{2.33}), with the parameter $\mu_2=\mu\neq0$ in both cases.

If the functions $f_{ij}$ are specified by (\ref{2.31}) with the parameter $\mu_2=\mu\neq0$, we derive
\begin{equation}\label{3.30}
\Delta_{13}=\mp\frac{\mu\eta_2}{\sqrt{1+\mu^2}}\phi_{1},\quad \Delta_{23}=\pm\frac{\eta_2}{\sqrt{1+\mu^2}}\phi_{1}.
\end{equation}
In light of (\ref{3.28}), we obtain $a=\pm1/\mu$. Substituting this expression into (\ref{3.26}) and (\ref{3.27}) yields
\begin{equation}\label{3.31}
\begin{pmatrix}
  -2b & a-c \\
  a-c & 2b
\end{pmatrix}
\begin{pmatrix}
  \Delta_{13} \\
  \Delta_{23}
\end{pmatrix}
=
\begin{pmatrix}
  0 \\
  0
\end{pmatrix}
\end{equation}
This gives $b=0$ and $a-c=0$, in contradiction with the Gauss equation (\ref{2.14}).

If the functions $f_{ij}$ are specified by (\ref{2.33}) with the parameter $\mu_2=\mu\neq0$, we get
\begin{equation}\label{3.32}
\Delta_{13}=\frac{2}{\gamma}\lambda\eta_2\eta_3 u_0 u_1,\quad \Delta_{23}=\frac{2}{\gamma}\lambda\eta_2(\mu\eta_3-\mu_3\eta_2)u_0 u_1.
\end{equation}
In views of (\ref{3.28}), we find
\begin{equation}\label{3.33}
\frac{\eta_2}{\gamma}\eta_3 u_0 u_1 a+\frac{\eta_2}{\gamma}(\mu\eta_3-\mu_3\eta_2)u_0 u_1\left(-\frac{a}{\mu}+2\right)=0.
\end{equation}
For $\eta_2(\mu\eta_3-\mu_3\eta_2)/\gamma\neq 0$ (equivalently, $\mu\eta_3-\mu_3\eta_2\neq0$), the equation (\ref{3.33}) immediately implies that $a$ must be constant. However, examining (\ref{3.26}) and (\ref{3.27}) reveals a contradiction identical to the previous case. For $\eta_2(\mu\eta_3-\mu_3\eta_2)/\gamma=0$, we obtain $\Delta_{13}=-2\lambda \eta_2 u_0 u_1$ and $\Delta_{23}=0$. Substituting these into (\ref{3.26}) and (\ref{3.27}) gives
\begin{equation}\label{3.34}
\begin{split}
&D_t a+\lambda u_0^2 D_xa-4\lambda(\pm\mu-a)u_0 u_1=0,\\
&D_t a+\lambda u_0^2 D_xa-2\lambda(a\mu^2-a\pm2\mu)u_0 u_1=0.
\end{split}
\end{equation}
This implies $a=0$ which, by Lemma 3.1, again leads to a contradiction.

Therefore, when $f_{21}\neq0$ and (\ref{3.18}) holds on a non-empty open set, $a$, $b$ and $c$ are universal functions of $x$ and $t$, independent of $u_0$.
$\hfill\square$
\end{Proof}

\begin{Lemma}\label{lemma3.4}
Consider a partial differential equation of the form (\ref{2.21}) which describes pseudospherical surfaces, given by Theorems 2.2-2.5. Assume there exists a local isometric immersion of the pseudospherical surface, determined by a solution $u(x,t)$ of the equation (\ref{2.21}), for which the coefficients $a$, $b$, $c$ of the second fundamental form depend on $x$, $t$, $u_0, \dots, u_l$, $w_1, \dots, w_n$, $v_1,\dots, v_n$, $1 \leq l,n < \infty$. If $f_{21}\neq0$ and
\begin{equation}\label{3.35}
c+\left(\frac{f_{11}}{f_{21}}\right)^2 a+2\frac{f_{11}}{f_{21}}b= 0
\end{equation}
on a non-empty open set, then $a$, $b$, and $c$ are universal functions of $x$ and $t$, independent of $u_0$.
is satisfied, then the system of equations (\ref{2.14})-(\ref{2.16}) is inconsistent.
\end{Lemma}

\begin{Proof}
If (\ref{3.35}) holds, then it follows form the Gauss equation (\ref{2.14}) that
\begin{equation}\label{3.36}
  b=\pm 1-\frac{f_{11}}{f_{21}}a,\quad c=\left(\frac{f_{11}}{f_{21}}\right)^2 a \mp2 \frac{f_{11}}{f_{21}}.
\end{equation}
Moreover, the following identities hold
\begin{equation}\label{3.37}
\begin{aligned}
&D_t b=-\frac{f_{11}}{f_{21}}D_t a-aD_t \left(\frac{f_{11}}{f_{21}}\right), &D_t c&=\left(\frac{f_{11}}{f_{21}}\right)^2 D_t a+2 \left(\frac{f_{11}}{f_{21}}a\mp1\right)D_t \left(\frac{f_{11}}{f_{21}}\right),\\
&D_x b=-\frac{f_{11}}{f_{21}}D_x a-aD_x \left(\frac{f_{11}}{f_{21}}\right), &D_x c&=\left(\frac{f_{11}}{f_{21}}\right)^2 D_x a+2 \left(\frac{f_{11}}{f_{21}}a\mp1\right)D_x \left(\frac{f_{11}}{f_{21}}\right),
\end{aligned}
\end{equation}
Then  the equations (\ref{2.15}) and (\ref{2.16}) can be rewritten as
\begin{equation}\label{3.38}
-af_{21}D_t \left(\frac{f_{11}}{f_{21}}\right)+\frac{\Delta_{12}}{f_{21}}D_x a+af_{22}D_x \left(\frac{f_{11}}{f_{21}}\right)-2b\Delta_{13}+(a-c)\Delta_{23}=0,
\end{equation}
and
\begin{equation}\label{3.39}
\begin{split}
&(f_{11}a\mp2f_{21})D_t \left(\frac{f_{11}}{f_{21}}\right)-\frac{f_{11}}{f_{21}}\frac{\Delta_{12}}{f_{21}}D_x a-\frac{\Delta_{12}}{f_{21}}a D_x \left(\frac{f_{11}}{f_{21}}\right)\\
&-f_{22}\left[\frac{f_{11}}{f_{21}}aD_x \left(\frac{f_{11}}{f_{21}}\right)\mp2D_x \left(\frac{f_{11}}{f_{21}}\right)\right]+(a-c)\Delta_{13}+2b\Delta_{23}=0.
\end{split}
\end{equation}
Taking a linear combination of above two equations results in
\begin{equation}\label{3.40}
\begin{split}
&\mp2f_{21}\left[\left(\frac{f_{11}}{f_{21}}\right)_{u_0} u_{0,t}+\left(\frac{f_{11}}{f_{21}}\right)_{u_1} u_{2,t}\right]-\left(\frac{\Delta_{12}}{f_{21}}a\mp2f_{22}\right)\left[\left(\frac{f_{11}}{f_{21}}\right)_{u_0} u_1+\left(\frac{f_{11}}{f_{21}}\right)_{u_2} u_3\right]\\
&+\left(a-c-2b\frac{f_{11}}{f_{21}}\right)\Delta_{13}+\left[\frac{f_{11}}{f_{21}}(a-c)+2b\right]\Delta_{23}=0,
\end{split}
\end{equation}
From (\ref{1.9}) and (\ref{2.23}), we obtain
\begin{equation}\label{3.41}
\left(\frac{f_{11}}{f_{21}}\right)_{u_0}+\left(\frac{f_{11}}{f_{21}}\right)_{u_2}=0,
\end{equation}
reducing (\ref{3.40}) to
\begin{equation}\label{3.42}
\begin{split}
&\left[\mp2f_{21}G\mp2(\lambda u_0^2f_{21}+f_{22})u_3-\left(\frac{\Delta_{12}}{f_{21}}a\mp2f_{22}\right)u_1+\frac{\Delta_{12}}{f_{21}}au_3 \right]\left(\frac{f_{11}}{f_{21}}\right)_{u_0}\\
&+\left[1+\left(\frac{f_{11}}{f_{21}}\right)^2\right]\left[a\Delta_{13}+\left(-\frac{f_{11}}{f_{21}}a\pm2\right)\Delta_{23}\right]=0.
\end{split}
\end{equation}
Taking the $v_k$ and $w_j$, $1\leq k,j\leq n$, derivatives of (\ref{3.42}), we get
\begin{equation}\label{3.43}
Pa_{v_k}=0,\quad Pa_{w_j}=0,
\end{equation}
where
\begin{equation}\label{3.44}
P=(u_3-u_1)\frac{\Delta_{12}}{f_{21}}\left(\frac{f_{11}}{f_{21}}\right)_{u_0}+\left[1+\left(\frac{f_{11}}{f_{21}}\right)^2\right]\left(\Delta_{13}-\frac{f_{11}}{f_{21}}\Delta_{23}\right).
\end{equation}

We now prove that $P\neq0$ by contradiction. Suppose, to the contrary, that $P=0$. Differentiating $P$ with respect to $u_3$ yields $\left(f_{11}/f_{21}\right)_{u_0}=0$, which implies that $f_{11}/f_{21}$ is a nonzero constant. However, this situation can only occur in the special cases described in Theorem 2.4 and Theorem 2.5(i). A careful examination of both cases reveals that they necessarily lead to $f_{11}=0$, which contradicts the nonzero constant assumption. Therefore, we conclude that $P$ must be nonzero.

Returning to (\ref{3.43}) and noting that $P\neq0$, we deduce $a_{v_k}=a_{w_j}=0$, for $k,j=1,2,\dots, n$. This implies that $a$ can depend only on $x$,$t$,$u_0,\dots,u_l$. However, further differentiation of (\ref{3.42}) with respect to $u_l$, $l\geq4$ leads to $Pa_{u_l}=0$, where $P$ is defined in (\ref{3.44}). Since $P\neq0$, we conclude that $a$ must in fact be independent of $u_l$ for $l\geq4$. Additionally, differentiating (\ref{3.38}) with respect to $u_4$ gives $a_{u_3}=0$. Finally, taking the $u_3$ derivatives of (\ref{3.42}), we have
\begin{equation}\label{3.45}
\left[\mp2(\lambda u_0^2f_{21}+f_{22})+\frac{\Delta_{12}}{f_{21}}a\right]\left(\frac{f_{11}}{f_{21}}\right)_{u_0}=0.
\end{equation}

When $\left(f_{11}/f_{21}\right)_{u_0}=0$, this condition is satisfied in the classification branches described by Theorem 2.4 and Theorem 2.5(i), both cases requiring  $\eta_2=0$ and $\mu_2=\mu\neq0$. Under the assumption that $f_{11}/f_{21}=1/\mu$, where $\mu$ is a nonzero constant, we calculate $\Delta_{13}$ and $\Delta_{23}$ and substituting the results into (\ref{3.42}), which shows that $a$ must be constant. Then (\ref{3.38}) and (\ref{3.39}) simplify to
\begin{equation}\label{3.46}
-2b\Delta_{13}+(a-c)\Delta_{23}=0,\quad (a-c)\Delta_{13}+2b\Delta_{23}=0.
\end{equation}
This implies that $b=0$ and $a=c$, which contradicts the Gauss equation (\ref{2.14}).

When $\left(f_{11}/f_{21}\right)_{u_0}\neq0$, this condition occurs in the classification branches described by Theorems 2.2-2.5. From (\ref{3.45}), by setting $\mp2(\lambda u_0^2f_{21}+f_{22})+\Delta_{12}/f_{21}a=0$ and utilizing $f_{i2}=\lambda u_0^2 f_{i1}+\phi_{i}$, we obtain
\begin{equation}\label{3.47}
a=\pm2\frac{\phi_{2}f_{21}}{\Delta_{12}}.
\end{equation}
Consequently, the equation (\ref{3.42}) can be reformulated as
\begin{equation}\label{3.48}
\left(\mp2f_{21}G\mp2\lambda u_0^2 u_1f_{21}\right)\left(\frac{f_{11}}{f_{21}}\right)_{u_0}
\pm2f_{21}\phi_{3}\left[1+\left(\frac{f_{11}}{f_{21}}\right)^2\right]=0,
\end{equation}
leading us to conclude
\begin{equation}\label{3.49}
G=-\lambda u_0^2 u_1+\left[1+\left(\frac{f_{11}}{f_{21}}\right)^2\right]\frac{\phi_{3}}{H},\quad H=\left(\frac{f_{11}}{f_{21}}\right)_{u_0}.
\end{equation}

If the functions $G$ and $f_{ij}$ are given by (\ref{2.30}) and (\ref{2.31}) (as specified in Theorem 2.2), we drive from (\ref{3.49}) that $H=\eta_2 f' /f_{21}^2$ and obtain the following equation
\begin{equation}\label{3.50}
\phi_{1,u_0}u_1+\phi_{1,u_1} u_2\pm\frac{\eta_2}{\sqrt{1+\mu_2^2}}\phi_{1}=\pm\frac{\sqrt{1+\mu_2^2}}{\eta_2}(f_{11}^2+f_{21}^2)\phi_{1}.
\end{equation}
Differentiating (\ref{3.50}) with respect to $u_2$ reveals the existence of a function $R=R(u_0)$ satisfying
\begin{equation}\label{3.51}
\frac{\phi_{12,u_1}}{\phi_{1}}=\pm\frac{\sqrt{1+\mu_2^2}}{\eta_2}(f_{11}^2+f_{21}^2)_{u_2}=R,
\end{equation}
from which we deduce
\begin{equation}\label{3.52}
\begin{split}
&\phi_{1}=Te^{Ru_1}, \quad T\neq 0,\\
&\pm\frac{\sqrt{1+\mu_2^2}}{\eta_2}(f_{11}^2+f_{21}^2)=Ru_2+S,
\end{split}
\end{equation}
where $T=T(u_0)$ and $S=S(u_0)$ are differentiable functions. Differentiating the second equation of (\ref{3.52}) with respect to $u_0$ and combining with its $u_2$ derivative, while noting that $f_{i1,u_0}=-f_{i1,u_2}$ for $i=1,2,3$, we determine that $R=r$ and $S=-ru_0+s$, where $r$ and $s$ are constants with $r\neq0$. Indeed, the case $r=0$ leads to $R=0$ and $S=s$, but differentiation of (\ref{3.52}) with respect to $u_2$ would then imply $f'=0$, resulting in a contradiction.
Substituting (\ref{3.52}) into (\ref{3.50}) yields
\begin{equation}\label{3.53}
T' u_1+T\left(\pm \frac{\eta_2}{\sqrt{1+\mu_2^2}}+ru_0-s\right)=0,
\end{equation}
the $u_1$ derivative of (\ref{3.53}) requires $T'=0$ and the coefficient of $u_0$ in (\ref{3.53}) gives $T=0$. This contradicts our initial assumption that $T\neq 0$.

If the functions $G$ and $f_{ij}$ are given by (\ref{2.32}) and (\ref{2.33}) (as specified in Theorem 2.3), similarly, we have $H=\eta_2 f' /f_{21}^2$ and the following equation
\begin{equation}\label{3.54}
u_0 u_1f+\frac{\eta_2}{\gamma}[u_1^2+u_0 u_2+(\mu_3 \eta_2-\mu_2 \eta_3)u_0 u_1]=\frac{\mu_3}{\gamma}(f_{11}^2+f_{21}^2)u_0 u_1.
\end{equation}
The coefficient of $u_1^2$ in the above equation requires $\eta_2/\gamma=0$, which contradicts the condition $\eta_2\neq0$ appearing in Theorem 2.3.

If the functions $G$ and $f_{ij}$ are given by (\ref{2.34}) and (\ref{2.35}) under the assumption $\eta_2\neq0$ (as specified in Theorem 2.4), similarly, we see $H=\eta_2 f' /f_{21}^2$ and the following equation
\begin{equation}\label{3.55}
\begin{split}
&u_{1}\phi_{1,u_0}+u_{2}\phi_{1,u_1}\pm\frac{\eta_2}{\sqrt{1+\mu_{2}^2}}\phi_{1}-\left(2\lambda u_{0}u_{1}\pm\frac { \eta_2 }{\sqrt{1+\mu_{2}^2}}\lambda u_{0}^{2}\pm\frac{C_1}{\sqrt{1+\mu_{2}^2}}\right)f\\
&=\frac{1}{\eta_2}(f_{11}^2+f_{21}^2)\left[\pm\sqrt{1+\mu_2^2}\phi_{1}\pm\frac{\mu_2}{\sqrt{1+\mu_2^2}}(\lambda \eta_2 u_0^2+C_1)\right].
\end{split}
\end{equation}
Differentiating (\ref{3.55}) with respect to $u_0$ and $u_2$ and adding the results leads to
\begin{equation}\label{3.56}
\begin{split}
&u_{1}\phi_{1,u_0 u_0}+u_{2}\phi_{1,u_1 u_0}\pm\frac{\eta_2}{\sqrt{1+\mu_{2}^2}}\phi_{1,u_0}-\left(2\lambda u_{1}\pm\frac {2 \eta_2 }{\sqrt{1+\mu_{2}^2}}\lambda u_{0}\right)f+\phi_{1,u_1}\\
&=\frac{1}{\eta_2}(f_{11}^2+f_{21}^2)\left[\pm\sqrt{1+\mu_2^2}\phi_{1,u_0}\pm2\lambda \frac{\mu_2\eta_2}{\sqrt{1+\mu_2^2}} u_0\right].
\end{split}
\end{equation}
Repeating the process on (\ref{3.56}) produces
\begin{equation}\label{3.57}
\begin{split}
&u_{1}\phi_{1,u_0 u_0 u_0}+u_{2}\phi_{1,u_1 u_0 u_0}\pm\frac{\eta_2}{\sqrt{1+\mu_{2}^2}}\phi_{1,u_0 u_0}\mp 2\lambda \frac{ \eta_2 }{\sqrt{1+\mu_{2}^2}}f+2\phi_{1,u_1u_0}\\
&=\frac{1}{\eta_2}(f_{11}^2+f_{21}^2)\left[\pm\sqrt{1+\mu_2^2}\phi_{1,u_0 u_0}\pm2\lambda \frac{\mu_2\eta_2}{\sqrt{1+\mu_2^2}}\right].
\end{split}
\end{equation}
A final application of the procedure to (\ref{3.57}) gives
\begin{equation}\label{3.58}
\begin{split}
&u_{1}\phi_{1,u_0 u_0 u_0 u_0}+u_{2}\phi_{1,u_1 u_0 u_0 u_0}\pm\frac{\eta_2}{\sqrt{1+\mu_{2}^2}}\phi_{1,u_0 u_0 u_0}+3\phi_{1,u_1 u_0 u_0}\\
&=\pm\frac{\sqrt{1+\mu_2^2}}{\eta_2}(f_{11}^2+f_{21}^2)\phi_{1,u_0 u_0 u_0}.
\end{split}
\end{equation}
By taking the $u_2$ derivative of (\ref{3.58}), we obtain
\begin{equation}\label{3.59}
\phi_{1,u_1 u_0 u_0 u_0}=\pm\frac{2\sqrt{1+\mu_2^2}}{\eta_2}(f_{11}f_{11,u_2}+f_{21}f_{21,u_2})\phi_{1,u_0 u_0 u_0}.
\end{equation}
This result naturally leads us to consider two distinct cases: $\phi_{1,u_0 u_0 u_0}=0$ and $\phi_{1,u_0 u_0 u_0}\neq0$.

Suppose $\phi_{1,u_0 u_0 u_0}=0$. From (\ref{3.59}), we find $\phi_{1,u_1 u_0 u_0 u_0}=0$ and (\ref{3.58}) subsequently yields $\phi_{1,u_1 u_0 u_0}=0$. Hence, $\phi_{1}=Au_0^2+Bu_0+D$, where $A\in\mathbb{R}$ and $B=B(u_1)$ and $D=D(u_1)$ are differentiable functions. Substitution into (\ref{3.57}) implies
\begin{equation}\label{3.60}
\pm2A\frac{\eta_2}{\sqrt{1+\mu_{2}^2}}\mp 2\lambda \frac{ \eta_2 }{\sqrt{1+\mu_{2}^2}}f+2B'
=\frac{1}{\eta_2}(f_{11}^2+f_{21}^2)\left[\pm2A\sqrt{1+\mu_2^2}\pm2\lambda\frac{\mu_2\eta_2}{\sqrt{1+\mu_2^2}}\right].
\end{equation}
First $u_2$ derivative of (\ref{3.60}), using $f'\neq0$, leads to
\begin{equation}\label{3.61}
\mp \frac{ \lambda \eta_2 }{\sqrt{1+\mu_{2}^2}}=-\frac{2}{\eta_2}(f_{11}+\mu_2f_{21})\left[\pm A\sqrt{1+\mu_2^2}\pm\frac{\lambda\mu_2\eta_2}{\sqrt{1+\mu_2^2}}\right].
\end{equation}
Second $u_2$ derivative yields
\begin{equation}\label{3.62}
\pm A\sqrt{1+\mu_2^2}\pm\frac{\lambda\mu_2\eta_2}{\sqrt{1+\mu_2^2}}=0.
\end{equation}
This enforces $\lambda=0$ and consequently $A=0$. Furthermore, (\ref{3.60}) implies $B$ must be constant. With these simplifications,  the equation (\ref{3.56}), reduces to
\begin{equation}\label{3.63}
\pm\frac{\eta_2}{\sqrt{1+\mu_2^2}}B+D'=\pm\frac{ \sqrt{1+\mu_2^2}}{\eta_2}(f_{11}^2+f_{21}^2)B.
\end{equation}
Double differentiation with respect to $u_2$ forces $B=0$, and thus $D'=0$. The equation (\ref{3.55}) simplifies to
\begin{equation}\label{3.64}
\pm\frac{\eta_2}{\sqrt{1+\mu_2^2}}D\mp\frac{C_1}{\sqrt{1+\mu_2^2}}f=\frac{ 1}{\eta_2}(f_{11}^2+f_{21}^2)\left[\pm\sqrt{1+\mu_2^2}D\pm\frac{\mu_2 C_1}{\sqrt{1+\mu_2^2}}\right].
\end{equation}
Two $u_2$ derivatives of (\ref{3.64}) require $C_1=0$, which contradicts the condition $(\lambda\eta_2)^2+C_1^2\neq0$ in Theorem 2.4 when combined with $\lambda=0$.

Suppose $\phi_{1,u_0 u_0 u_0}\neq0$. It follows from (\ref{3.59}) that
\begin{equation}\label{3.65}
\frac{\phi_{1,u_1 u_0 u_0 u_0}}{\phi_{1,u_0 u_0 u_0}}=\pm\frac{\sqrt{1+\mu_2^2}}{\eta_2}(f_{11}^2+f_{21}^2)_{u_2}=R,
\end{equation}
where $R=R(u_0)$ is a differentiable function. This decouples into two equations
\begin{equation}\label{3.66}
\begin{split}
&\phi_{1,u_1 u_0 u_0 u_0}=R\phi_{1,u_0 u_0 u_0}, \\
& f_{11}^2+f_{21}^2=\pm\frac{\eta_2}{\sqrt{1+\mu_2^2}}Ru_2 +S,
\end{split}
\end{equation}
where $S=S(u_0)$ also is a differentiable function.  Applying the condition $f_{i1,u_0}+f_{i1,u_2}=0$ to the second equation of (\ref{3.66}) yields
\begin{equation}\label{3.67}
\begin{split}
&R=A,\\
&S=\mp\frac{\eta_2}{\sqrt{1+\mu_2^2}}Au_0+B,
\end{split}
\end{equation}
with $A$ and $B$ two constants. Consequently, we obtain
\begin{equation}\label{3.68}
f_{11}^2+f_{21}^2=\mp\frac{\eta_2}{\sqrt{1+\mu_2^2}}A(u_0-u_2) +B,
\end{equation}
and through integration of the first equation of (\ref{3.66}), we have
\begin{equation}\label{3.69}
\phi_{1,u_1 u_0 u_0}=A\phi_{1,u_0 u_0}+D,
\end{equation}
where $D=D(u_1)$ is a differentiable function. Substituting into (\ref{3.57}) leads to
\begin{equation}\label{3.70}
\begin{split}
&u_{1}\phi_{1,u_0 u_0 u_0}+u_{2}(A\phi_{1,u_0 u_0}+D)\pm\frac{\eta_2}{\sqrt{1+\mu_{2}^2}}\phi_{1,u_0 u_0}\mp \frac{2\lambda \eta_2 }{\sqrt{1+\mu_{2}^2}}f+2\phi_{1,u_1u_0}\\
&=\left(\mp\frac{A(u_0 -u_2)}{\sqrt{1+\mu_2^2}}+B\right)\left[\pm\sqrt{1+\mu_2^2}\phi_{1,u_0 u_0}\pm \frac{2\lambda\mu_2\eta_2}{\sqrt{1+\mu_2^2}}\right].
\end{split}
\end{equation}
Taking the $u_2$ derivative of (\ref{3.70}) returns
\begin{equation}\label{3.71}
\pm \frac{2\lambda \eta_2 }{\sqrt{1+\mu_{2}^2}}f'=-D+ \frac{2\lambda\mu_2\eta_2}{\sqrt{1+\mu_2^2}}A\equiv \pm \frac{2\lambda\eta_2}{\sqrt{1+\mu_2^2}}E,
\end{equation}
with $E$ a nonzero constant, due to $f'\neq0$. This implies $f_{11}=f=E(u_0-u_2)+F$, where $F$ is a constant. On the other hand, using $f_{21}=\mu_2 f_{11}+\eta_2$ in (\ref{3.68}) gives
\begin{equation}\label{3.72}
(1+\mu_2^2)f_{11}^2+2\mu_2\eta_2f_{11}+\eta_2^2=\mp\frac{\eta_2}{\sqrt{1+\mu_2^2}}A(u_0-u_2) +B.
\end{equation}
Double differentiation of $u_0$ combined with the linear form of $f_{11}$ produces $E=0$, which is a contradiction with $f'\neq0$.

Therefore, we show that neither $\phi_{1,u_0 u_0 u_0}=0$ nor $\phi_{1,u_0 u_0 u_0}\neq0$ yields consistent solutions, thereby invalidating (\ref{3.49}) is not true if the functions $G$ and $f_{ij}$ are given by Theorem 2.4 under the assumption $\eta_2\neq 0$.

If the functions $G$ and $f_{ij}$ are given by (\ref{2.36}) and (\ref{2.37}) under the assumption $\eta_2\neq0$ (as specified in Theorem 2.5 (i)), similarly, we find $H=\nu\eta_2 /f_{21}^2$ and the following equation
\begin{equation}\label{3.73}
\begin{split}
&\lambda \left[-5u_{0}^{2}u_{1}+4u_{0}u_{1}u_{2}+\left(2\zeta_{1}-\frac{4}{\theta}\right)u_{0}u_{1}+\frac{2\zeta_{1}}{\theta}u_{1}-\frac{2}{\theta}u_{1}u_{2}\right]\\
&+\left[\theta u_{1}^{3}+2u_{0}u_{1}+u_{1}u_{2}-\nu_{1}u_{1}\right]\theta C_2e^{\theta u_0}=-\lambda u_0^2 u_1
+(f_{11}^2+f_{21}^2)\frac{\phi_{3}}{\nu\eta_2},
\end{split}
\end{equation}
with
\begin{equation}\label{3.74}
\phi_{3}=-\frac{\mu_3 \nu}{\theta}(2\lambda-\theta^2 C_2e^{\theta u_0})u_1^2+\left(\frac{2\lambda}{\theta}-\theta C_2e^{\theta u_0}+2\lambda u_0\right)\left(\mu_2 u_1-\frac{\eta_3}{\theta}\right).
\end{equation}
Triple differentiation of (\ref{3.73}) with respect to $u_1$ yields $C_2=0$. Consequently, we confirm $\lambda\neq0$. With $C_2$ eliminated, (3.73) reduces to
\begin{equation}\label{3.75}
\begin{split}
&\lambda \left[-4u_{0}^{2}u_{1}+4u_{0}u_{1}u_{2}+\left(2\zeta_{1}-\frac{4}{\theta}\right)u_{0}u_{1}+\frac{2\zeta_{1}}{\theta}u_{1}-\frac{2}{\theta}u_{1}u_{2}\right]\\
&=(f_{11}^2+f_{21}^2)\frac{1}{\nu\eta_2}\left[-2\lambda\frac{ \mu_3 \nu}{\theta} u_1^2 +\left(\frac{2\lambda}{\theta}+2\lambda u_0\right)\left(\mu_2 u_1-\frac{\eta_3}{\theta}\right)\right].
\end{split}
\end{equation}
Double differentiation of (\ref{3.75}) with respect to $u_1$ gives
\begin{equation}\label{3.76}
(f_{11}^2+f_{21}^2)\frac{1}{\nu\eta_2}\left(-2\lambda\frac{ \mu_3 \nu}{\theta}\right)=0,
\end{equation}
implying $\mu_3=0$. However, this directly contradicts the condition $\mu_3=\pm\sqrt{1+\mu_2^2}\neq 0$.

If the functions $G$ and $f_{ij}$ are given by (\ref{2.38}) and (\ref{2.39}) (as specified in Theorem 2.5 (ii)), similarly, we get $H=\nu\eta_2 /f_{21}^2$ and the following equation
\begin{equation}\label{3.77}
\begin{split}
&\lambda \left[-3u_{0}^{2}u_{1}+2u_{0}u_{1}u_{2}+2\zeta_{2}u_{0}u_{1}\mp \frac{2}{\tau}(u_{1}^{2}+u_{0}u_{2})\right]
+\tau \left(\pm u_{1}+\tau u_{0}u_{2}-\zeta_{2}\tau u_{2}\right)\varphi e^{\pm \tau u_1}
\\&\pm(\tau u_{0}u_{1}\pm u_{2}+\tau u_{1}u_{2}-\zeta_{2}\tau u_{1})\varphi^{'}e^{\pm \tau u_1}+\varphi^{''}u_{1}^{2}e^{\pm \tau u_1}
=-\lambda u_0^2 u_1+(f_{11}^2+f_{21}^2)\frac{\phi_{3}}{\nu\eta_2},
\end{split}
\end{equation}
with
\begin{equation}\label{3.78}
 \phi_{3}=\mu_3 [\pm\tau(\nu u_{0}-\sigma)\varphi+\nu\varphi 'u_{1}]e^{\pm \tau u_1}\mp \frac{2\lambda \nu\mu_3}{\tau}u_{0}u_{1}\pm\eta_3\tau\varphi e^{\pm \tau u_1}.
\end{equation}
Taking the $u_2$ derivative twice of (\ref{3.77}) returns
\begin{equation}\label{3.79}
-\frac{2\nu}{\eta_2}(1+\mu_2^2)=0.
\end{equation}
This implies $\nu=0$, which is a contradiction.

Therefore, we demonstrate that the expression (\ref{3.49}) cannot hold when the functions $G$ and $f_{ij}$ are specified according to any of the cases presented in Theorem 2.2-2.5. This concludes the proof of Lemma 3.4.
$\hfill\square$
\end{Proof}

Thus far we demonstrated that if there exists a local isometric immersion of the pseudospherical surface, determined by a solution $u(x,t)$ of the equation (\ref{2.21}), for which the coefficients $a$, $b$, $c$ of the second fundamental form depend on $x$, $t$, $u_0, \dots, u_l$, $w_1, \dots, w_m$, $v_1,\dots, v_n$, $1 \leq l,m,n < \infty$, then $a$, $b$, and $c$ are universal functions of $x$ and $t$, independent of $u_0$. Building upon this result, we proceed to explicitly determine these coefficients for all cases presented in Theorems 2.2-2.5.

\begin{Proposition}\label{proposition3.5}
Consider a partial differential equation of the form (\ref{2.21}) which describes pseudospherical surfaces, with $f_{ij}$ given by Theorems 2.2. There is a local isometric immersion of the pseudospherical surface, determined by a solution $u(x,t)$ of the equation (\ref{2.21}), for which the coefficients $a$, $b$, $c$ of the second fundamental form depend on $x$, $t$, $u_0, \dots, u_l$, $w_1, \dots, w_m$, $v_1,\dots, v_n$, $1 \leq l,m,n < \infty$, if and only if $a$, $b$, $c$ depend only on $x$ and are given by

\noindent$(i)$ When $ \mu_2=0$,
\begin{equation}\label{3.80}
a=\pm \sqrt{L(x)}, \quad b=-\beta e^{\pm2\eta_2x}, \quad c=a\mp\frac{a_x}{\eta_2},
\end{equation}
where $L(x)=\alpha e^{\pm2\eta_2 x}-\beta^2 e^{\pm4\eta_2 x}-1$, with real constants $\alpha,\beta$ satisfying $\alpha>0$ and $\alpha^2>4\beta^2$. The 1-forms $\omega_i$ are defined on the strip
\begin{equation}\label{3.81}
 \frac{\alpha-\sqrt{\alpha^2-4\beta^2}}{2\beta^2}<e^{\pm2\eta_2 x}< \frac{\alpha+\sqrt{\alpha^2-4\beta^2}}{2\beta^2},
\end{equation}
where $\alpha$ and $\beta$ must be chosen such that the strip intersects the domain of the solution of (\ref{2.30}).

\noindent$(ii)$ When $ \mu_2 \neq0$,
\begin{equation}\label{3.82}
  \begin{aligned}
    a&= \frac{1}{2\mu_2}\left[\pm\mu_2\sqrt{\Delta}+(1-\mu_2^2)b+\beta e^{\pm\frac{2\eta_2x}{\sqrt{1+\mu_2^2}}}\right],\\
    c&= \frac{1}{2\mu_2}\left[\pm\mu_2\sqrt{\Delta}-(1-\mu_2^2 )b-\beta e^{\pm\frac{2\eta_2x}{\sqrt{1+\mu_2^2}}}\right],\\
    \Delta &=\frac{\left[(\mu_2^2-1)b-\beta e^{\pm\frac{2\eta_2x}{\sqrt{1+\mu_2^2}}}\right]^2-4\mu_2^2(1-b^2)}{\mu_2^2}>0,
  \end{aligned}
\end{equation}
where $b$ satisfies the nonlinear ordinary differential equation
\begin{equation}\label{3.83}
\begin{split}
 & \left[\pm(\mu_2^2+1)^2 b\mp(\mu_2^2 -1)\beta e^{\pm\frac{2\eta_2x}{\sqrt{1+\mu_2^2}}}+\mu_2(\mu_2^2 +1)\sqrt{\Delta}\right]b_x \\
 & +\frac{2\eta_2}{\sqrt{1+\mu_2^2}}\left[\mp\mu_2(\mu_2^2+1)\sqrt{\Delta}b-(\mu_2^2-1)\beta e^{\pm\frac{2\eta_2x}{\sqrt{1+\mu_2^2}}}b+\beta^2 e^{\pm\frac{4\eta_2x}{\sqrt{1+\mu_2^2}}}\right]=0.
\end{split}
\end{equation}
\end{Proposition}

\begin{Proof}
The non-vanishing condition $\eta_2\neq0$ guarantees that $f_{21}\neq  0$ on an open set. By Lemma 3.4, this implies the inconsistency of the system (\ref{2.14})-(\ref{2.16}). Consequently, Lemma 3.3 holds, yielding the following forms of (\ref{2.15}) and (\ref{2.16})
\begin{equation}\label{3.84}
f a_t +(\mu_2 f+\eta_2)b_t-\phi_{1}a_x-\mu_2 \phi_{1}b_x\pm2b\frac{\mu_2 \eta_2}{\sqrt{1+\mu_2^2}}\phi_{1}\pm(a-c)\frac{\eta_2}{\sqrt{1+\mu_2^2}}\phi_{1}=0,
\end{equation}
\begin{equation}\label{3.85}
f b_t +(\mu_2 f+\eta_2)c_t-\phi_{1}b_x-\mu_2 \phi_{1}c_x\mp(a-c)\frac{\mu_2 \eta_2}{\sqrt{1+\mu_2^2}}\phi_{1}\pm2b\frac{\eta_2}{\sqrt{1+\mu_2^2}}\phi_{1}=0.
\end{equation}
Differentiation of (\ref{3.84}) and (\ref{3.85}) with respect to $u_2$ leads to
\begin{equation}\label{3.86}
a_t=\mu_2^2 c_t,\quad b_t=-\mu_2 c_t.
\end{equation}
Substituting these into (\ref{3.84}) and (\ref{3.85}) gives
\begin{equation}\label{3.87}
-\mu_2 \eta_2 c_t+\phi_{1}\left[-a_x-\mu_2 b_x\pm\frac{\eta_2}{\sqrt{1+\mu_2^2}}(2\mu_2 b+a-c)\right]=0,
\end{equation}
\begin{equation}\label{3.88}
\eta_2 c_t+\phi_{1}\left[-b_x-\mu_2 c_x\pm\frac{\eta_2}{\sqrt{1+\mu_2^2}}(2b-\mu_2 a+\mu_2 c)\right]=0.
\end{equation}
The linear combination of (\ref{3.87}) and (\ref{3.88}) eliminates $c_t$, producing
\begin{equation}\label{3.89}
\mu_2 \left[-b_x-\mu_2 c_x\pm\frac{\eta_2}{\sqrt{1+\mu_2^2}}(2b-\mu_2 a+\mu_2 c)\right]
+\left[-a_x-\mu_2 b_x\pm\frac{\eta_2}{\sqrt{1+\mu_2^2}}(2\mu_2 b+a-c)\right]=0.
\end{equation}
Differentiating (\ref{3.89}) with respect to $t$ and using (\ref{3.86}) reveals $ a_t=b_t=c_t=0$. Hence, $a$, $b$ and $c$ only depend on $x$ and (\ref{3.84}) and (\ref{3.85}) become
\begin{equation}\label{3.90}
-a_x-\mu_2 b_x\pm\frac{\eta_2}{\sqrt{1+\mu_2^2}}(2\mu_2 b+a-c)=0,
\end{equation}
\begin{equation}\label{3.91}
-b_x-\mu_2 c_x\pm\frac{\eta_2}{\sqrt{1+\mu_2^2}}(2b-\mu_2 a+\mu_2 c)=0.
\end{equation}
Solving (\ref{3.90}) for $c$ in terms of $a$, $b$, and their derivatives, and substituting into (\ref{3.91}), we obtain
\begin{equation}\label{3.92}
\mu_2\left(\pm\frac{\sqrt{1+\mu_2^2}}{\eta_2}-2a_x\right)+\mu_2^2\left(\pm\frac{\sqrt{1+\mu_2^2}}{\eta_2}b_{xx}-2b_x\right)
-(1+\mu_2^2)\left(b_x\mp\frac{2\eta_2}{\sqrt{1+\mu_2^2}}b\right)=0,
\end{equation}
which is,
\begin{equation}\label{3.93}
\mu_2 a_x=-\mu_2^2 b_x\pm\eta_2\sqrt{1+\mu_2^2}b\pm\frac{\beta \eta_2}{\sqrt{1+\mu_2^2}}e^{\pm\frac{2 \eta_2x}{\sqrt{1+\mu_2^2}}},
\end{equation}
with $\beta$ an integration constant.

If $\mu_2=0$, from (\ref{3.90}) and (\ref{3.93}), we have
\begin{equation}\label{3.94}
b=-\beta e^{\pm2\eta_2 x}, \quad c=a\mp\frac{a_x}{\eta_2}.
\end{equation}
Substituting these expressions into the Gauss equation (\ref{2.14}), we derive
\begin{equation}\label{3.95}
a=\pm\sqrt{L(x)},\quad L(x)=\alpha e^{\pm2\eta_2 x}-\beta^2 e^{\pm4\eta_2 x}-1,
\end{equation}
with real constants $\alpha$, $\beta$ satisfying $\alpha>0$ and $\alpha^2>4\beta^2$, where $a$ is defined on the strip given by inequality (\ref{3.81}).

If $\mu_2\neq0$, combining (\ref{3.90}) and (\ref{3.93}) leads to
\begin{equation}\label{3.96}
c=a+\Phi, \quad \Phi=\Phi (x)=\frac{\mu_2^2-1}{\mu_2}b-\frac{\beta}{\mu_2}e^{\pm\frac{2\eta_2 x}{\sqrt{1+\mu_2^2}}}.
\end{equation}
Substituting into the Gauss equation (\ref{2.14}) produces a second order equation
\begin{equation}\label{3.97}
a^2+a\Phi-b^2+1=0,
\end{equation}
which determines $a$ and $c$ as given in (\ref{3.82}). The equation (\ref{3.93}) can be reformulated as
\begin{equation}\label{3.98}
b_x[(\mu_2^2+1)\sqrt{\Delta}\pm(\mu_2^2-1)\Phi\pm4\mu_2 b]\mp2\eta_2\sqrt{1+\mu_2^2}b\sqrt{\Delta}-\frac{2\beta\eta_2}{\sqrt{1+\mu_2^2}}\Phi e^{\pm\frac{2\eta_2 x}{\sqrt{1+\mu_2^2}}}=0.
\end{equation}
The vanishing of the coefficient of $b_x$ would require
\begin{align}
&(\mu_2^2+1)\sqrt{\Delta}\pm(\mu_2^2-1)\Phi\pm4\mu_2 b=0, \label{3.99}
\\&\mp2\eta_2\sqrt{1+\mu_2^2}b\sqrt{\Delta}-\frac{2\beta\eta_2}{\sqrt{1+\mu_2^2}}\Phi e^{\pm\frac{2\eta_2 x}{\sqrt{1+\mu_2^2}}}=0.\label{3.100}
\end{align}
However, these conditions lead to
\begin{equation}\label{3.101}
\frac{\mu_2\eta_2}{\sqrt{1+\mu_2^2}}(\Phi^2+4b^2)=0,
\end{equation}
which implies $\Phi=b=0$ and consequently $c=a$. This contradicts the Gauss equation (\ref{2.14}), proving the coefficient of $b_x$ in (\ref{3.98}) cannot vanish. Therefore, we can assume $b_x=g(x,b)$, where $g(x,b)$ is a differentiable function defined as
\begin{equation}\label{3.102}
g(x,b)=\frac{\pm2\eta_2\sqrt{1+\mu_2^2}b\sqrt{\Delta}+\frac{2\beta\eta_2}{\sqrt{1+\mu_2^2}}\Phi e^{\pm\frac{2\eta_2 x}{\sqrt{1+\mu_2^2}}}}{(\mu_2^2+1)\sqrt{\Delta}\pm(\mu_2^2-1)\Phi\pm4\mu_2 b}.
\end{equation}

Consider a fixed point $x_0$ and the associated initial value problem
\begin{equation}\label{3.103}
\begin{cases}
b_x=g(x,b),\\
 b(x_0)=b_0,
 \end{cases}
\end{equation}
where $g(x,b)$ is given by (\ref{3.102}). The function $g(x,b)$ and its partial derivative with respect to $b$ are both continuous in an open rectangular domain
\begin{equation}\label{3.104}
R=\{(x,b):x_1<x<x_2,b_1<b<b_2\}
\end{equation}
containing the initial point $(x_0,b_0)$. By the fundamental existence and uniqueness theorem for ordinary differential equations, there exists unique solution $b(x)$ on the interval $I=[b_0-\epsilon,b_0+\epsilon]$, where $\epsilon$ is a positive constant. Additionally, the interval boundaries $x_1$, $x_2$ must be chosen such that $(x_1,x_2)$ intersects with the domain of the solution of (\ref{2.30}). Finally, substituting the expression for $\Phi$ into (\ref{3.98}) produces (\ref{3.83}).

The converse statement follows directly through straightforward computation.
$\hfill\square$
\end{Proof}

\begin{Proposition}\label{proposition3.6}
Consider a partial differential equation of the form (\ref{2.21}) which describes pseudospherical surfaces, with $f_{ij}$ given by Theorems 2.3. There is no local isometric immersion of the pseudospherical surface, determined by a solution $u(x,t)$ of the equation (\ref{2.21}), for which the coefficients $a$, $b$, $c$ of the second fundamental form depend on $x$, $t$, $u_0, \dots, u_l$, $w_1, \dots, w_m$, $v_1,\dots, v_n$, $1 \leq l,m,n < \infty$.
\end{Proposition}

\begin{Proof}
The non-vanishing condition $\eta_2\neq0$ guarantees that $f_{21}\neq  0$ on an open set. Lemma 3.4 shows that the system of equations (\ref{2.14})-(\ref{2.16}) is inconsistent. Consequently, Lemma 3.3 holds and (\ref{2.15}) and (\ref{2.16}) reduce to
\begin{equation}\label{3.105}
\begin{split}
&f[a_t+\mu_2 b_t+\lambda(a_x+\mu_2 b_x)u_0^2]+\eta_2(b_t+\lambda u_0^2 b_x)\\
&+\lambda u_0 u_1\left[\frac{2}{\gamma}\eta_2(a_x+\mu_2 b_x)-\frac{4}{\gamma}\eta_2\eta_3 b-\frac{2}{\gamma}\eta_2(\mu_3\eta_2-\mu_2\eta_3)(a-c)\right]=0,
\end{split}
\end{equation}
\begin{equation}\label{3.106}
\begin{split}
&f[b_t+\mu_2 c_t+\lambda(b_x+\mu_2 c_x)u_0^2]+\eta_2(c_t+\lambda u_0^2 c_x)\\
&+\lambda u_0 u_1\left[\frac{2}{\gamma}\eta_2(b_x+\mu_2 c_x)+\frac{2}{\gamma}\eta_2\eta_3 (a-c)-\frac{4}{\gamma}\eta_2(\mu_3\eta_2-\mu_2\eta_3)b\right]=0.
\end{split}
\end{equation}
Differentiating (\ref{3.105}) and (\ref{3.106}) with respect to $u_2$ leads to
\begin{equation}\label{3.107}
a_t+\mu_2 b_t+\lambda(a_x+\mu_2 b_x)u_0^2=0,
\end{equation}
\begin{equation}\label{3.108}
b_t+\mu_2 c_t+\lambda(b_x+\mu_2 c_x)u_0^2=0.
\end{equation}
By comparing coefficients of $u_0^2$, we have
\begin{equation}\label{3.10}
a_t+\mu_2 b_t=0,\quad a_x+\mu_2 b_x=0,
\end{equation}
\begin{equation}\label{3.110}
b_t+\mu_2 c_t=0,\quad b_x+\mu_2 c_x=0.
\end{equation}
Substituting these back into (\ref{3.105}) and (\ref{3.106}) and differentiating successively with respect to $u_1$ and $u_0$ gives
\begin{equation}\label{3.111}
-\frac{4}{\gamma}\eta_2\eta_3 b-\frac{2}{\gamma}\eta_2(\mu_3\eta_2-\mu_2\eta_3)(a-c)=0,
\end{equation}
\begin{equation}\label{3.112}
\frac{2}{\gamma}\eta_2\eta_3 (a-c)-\frac{4}{\gamma}\eta_2(\mu_3\eta_2-\mu_2\eta_3)b=0.
\end{equation}
The Gauss equation (\ref{2.14}) requires $(a-c)^2+b^2\neq0$. However, from (\ref{3.111}) and (\ref{3.112}), we obtain $\eta_2^2\eta_3^2+\eta_2^2(\mu_3\eta_2-\mu_2\eta_3)^2=0$, implying $\eta_2=0$, which is a contradiction.
$\hfill\square$
\end{Proof}

\begin{Proposition}\label{proposition3.7}
Consider a partial differential equation of the form (\ref{2.21}) which describes pseudospherical surfaces, with $f_{ij}$ given by Theorems 2.4. There is a local isometric immersion of the pseudospherical surface, determined by a solution $u(x,t)$ of the equation (\ref{2.21}), for which the coefficients $a$, $b$, $c$ of the second fundamental form depend on $x$, $t$, $u_0, \dots, u_l$, $w_1, \dots, w_m$, $v_1,\dots, v_n$, $1 \leq l,m,n < \infty$, if and only if $a$, $b$, $c$ depend only on $x$ and $t$ and are given by

\noindent$(i)$ When $ \mu_2=\eta_2=0$, $C_1\neq0$,
\begin{equation}\label{3.113}
a=\pm \sqrt{L(t)}, \quad b=\beta e^{\pm2C_1t}, \quad c=a\mp\frac{a_t}{C_1},
\end{equation}
where $L(t)=\alpha e^{\pm2C_1t}-\beta^2 e^{\pm4C_1t}-1$, with real constants $\alpha,\beta$ satisfying $\alpha>0$ and $\alpha^2>4\beta^2$. The 1-forms $\omega_i$ are defined on the strip
\begin{equation}\label{3.114}
 \frac{\alpha-\sqrt{\alpha^2-4\beta^2}}{2\beta^2}<e^{\pm2C_1t}< \frac{\alpha+\sqrt{\alpha^2-4\beta^2}}{2\beta^2},
\end{equation}
where $\alpha$ and $\beta$ must be chosen such that the strip intersects the domain of the solution of (\ref{2.34}).

\noindent$(ii)$ When $\mu_2=0$, $\eta_2\neq0$, $\lambda^2+C^2\neq0$,
\begin{equation}\label{3.115}
a=\pm \sqrt{L(\xi)}, \quad b=-\beta e^{\pm2\xi}, \quad c=a\mp a_{\xi},
\end{equation}
where $L(\xi)=\alpha e^{\pm2\xi}-\beta^2 e^{\pm4\xi}-1,\xi=\eta_2 x+C_1t$, with real constants $\alpha,\beta$ satisfying $\alpha>0$ and $\alpha^2>4\beta^2$. The 1-forms $\omega_i$ are defined on the strip
\begin{equation}\label{3.116}
 \frac{\alpha-\sqrt{\alpha^2-4\beta^2}}{2\beta^2}<e^{\pm2\xi}< \frac{\alpha+\sqrt{\alpha^2-4\beta^2}}{2\beta^2},
\end{equation}
where $\alpha$ and $\beta$ must be chosen such that the strip intersects the domain of the solution of (\ref{2.34}).

\noindent$(iii)$ When $\mu_2 \neq0$, $(\lambda\eta_2)^2+C_1^2\neq0$,
\begin{equation}\label{3.117}
  \begin{aligned}
    a&= \frac{1}{2\mu_2}\left[\pm\mu_2\sqrt{\Delta}-(\mu_2^2 -1)b+\beta e^{\pm\frac{2\xi}{\sqrt{1+\mu_2^2}}}\right],\\
    c&= \frac{1}{2\mu_2}\left[\pm\mu_2\sqrt{\Delta}+(\mu_2^2 -1)b-\beta e^{\pm\frac{2\xi}{\sqrt{1+\mu_2^2}}}\right],\\
    \Delta &=\frac{\left[(\mu_2^2-1)b-\beta e^{\pm\frac{2\xi}{\sqrt{1+\mu_2^2}}}\right]^2-4\mu_2^2(1-b^2)}{\mu_2^2}>0,
  \end{aligned}
\end{equation}
where $\xi=\eta_2 x+C_1t$ and $b$ satisfies the nonlinear ordinary differential equation
\begin{equation}\label{3.118}
\begin{split}
 & \left[\pm(\mu_2^2+1)^2 b\mp(\mu_2^2 -1)\beta e^{\pm\frac{2\xi}{\sqrt{1+\mu_2^2}}}+\mu_2(\mu_2^2 +1)\sqrt{\Delta}\right]b_{\xi} \\
 & +\frac{2}{\sqrt{1+\mu_2^2}}\left[\mp\mu_2(\mu_2^2+1)\sqrt{\Delta}b-(\mu_2^2-1)\beta e^{\pm\frac{2\xi}{\sqrt{1+\mu_2^2}}}b+\beta^2 e^{\pm\frac{4\xi}{\sqrt{1+\mu_2^2}}}\right]=0.
\end{split}
\end{equation}
\end{Proposition}

\begin{Proof}
Suppose $f_{21}=0$, which implies $\mu_2=\eta_2=0$ and $C_1\neq0$. Consequently, Lemma 3.2 holds and (\ref{2.15}) and (\ref{2.16}) become
\begin{align}
&f[a_t+\lambda u_0^2 a_x\mp C_1(a-c)]-\phi_{1}a_x-C_1b_x=0,\label{3.119}
\\&f[b_t+\lambda u_0^2 b_x\mp2bC_1]-\phi_{1}b_x-C_1c_x=0.\label{3.120}
\end{align}
Differentiation of (\ref{3.119}) and (\ref{3.120}) with respect to $u_2$ leads to
\begin{align}\label{3.121}
&a_t+\lambda u_0^2 a_x\mp C_1(a-c)=0,\\
\label{3.122}
&b_t+\lambda u_0^2 b_x\mp2bC_1=0.
\end{align}
By comparing the $u_0^2$ coefficients, we obtain
\begin{align}
&\lambda a_x=0,\quad a_t\mp C_1(a-c)=0,\label{3.123}
\\&\lambda b_x=0,\quad b_t\mp2bC=0.\label{3.124}
\end{align}
Substituting beck into (\ref{3.119}) and (\ref{3.120}) gives
\begin{equation}\label{3.125}
b_x=-\frac{\phi_{1}}{C_1}a_x,\quad c_x=-\frac{\phi_{1}}{C_1}b_x.
\end{equation}
Differentiating the Gauss equation (\ref{2.14}) with respect to $x$ and applying (\ref{3.125}) returns
\begin{equation}\label{3.126}
\left[c+\left(\frac{\phi_{1}}{C_1}\right)^2a-2\frac{\phi_{1}}{C_1}b\right]a_x=0.
\end{equation}

If $a_x\neq0$, then the expression between brackets in (\ref{3.126}) vanishes. Differentiation of the first equation in (\ref{3.125}) with respect to $u_0$ and $u_1$ gives $\phi_{1,u_0}=\phi_{1,u_1}=0$, implying $\phi_{1}$ is a non-zero constant. Setting $\phi_{1}=C_1\alpha$ with $\alpha \in\mathbb{R} \setminus \{0\}$, the Gauss equation (\ref{2.14}) yields
\begin{equation}\label{3.127}
b=\pm1-\alpha a,\quad c=\alpha^2 a\mp2\alpha.
\end{equation}
Substituting into (\ref{3.123}) and (\ref{3.124}) leads to
\begin{equation}\label{3.128}
a_t\mp C_1(a-\alpha a\pm2\alpha)=0,\quad -\alpha a_t\mp2C_1(\pm1-\alpha a)=0.
\end{equation}
The linear combination of above two equations forces $a=\pm2/\alpha$, which when substituted back leads to $C_1=0$. This contradicts the assumption that $C_1\neq0$.

If $a_x=0$, from (\ref{3.125}), we obtain $b_x=c_x=0$. Thus, (\ref{3.123}) and (\ref{3.124}) admit solutions
\begin{equation}\label{3.129}
b=\beta e^{\pm2C_1t},\quad c=a\mp\frac{a_t}{C_1},
\end{equation}
where $\beta$ is an integration constant. Substituting into the Gauss equation (\ref{2.14}) leads to a second order equation
\begin{equation}\label{3.130}
a^2C_1\mp aa_t-\beta^2C_1e^{\pm4C_1t}+C_1=0,
\end{equation}
then we derive $a=\pm\sqrt{L(t)}$, $L(t)=\alpha e^{\pm2C_1t}-\beta^2 e^{\pm4C_1t}-1$, with real constants $\alpha$, $\beta$ satisfying $\alpha>0$ and $\alpha^2>4\beta^2$, where $a$ is defined on the strip given by (\ref{3.114}). This concludes (i).

Suppose $f_{21}\neq0$, from Lemma 3.4, the equations (\ref{2.14})-(\ref{2.16}) form an inconsistent system. Hence, Lemma 3.3 holds and (\ref{2.15}) and (\ref{2.16}) become
\begin{equation}\label{3.131}
\begin{split}
&f\left[a_t+\mu_2 b_t+\lambda(a_x+\mu_2 b_x)u_0^2\mp\frac{2b\mu_2}{\sqrt{1+\mu_2^2}}(C_1+\lambda \eta_2 u_0^2)\mp\frac{(a-c)}{\sqrt{1+\mu_2^2}}(C_1+\lambda\eta_2 u_0^2)\right]\\
&+(\eta_2b_t-C_1b_x)+\phi_{1}\left[-a_x-\mu_2b_x\pm\frac{2b\mu_2\eta_2}{\sqrt{1+\mu_2^2}}\pm\frac{\eta_2(a-c)}{\sqrt{1+\mu_2^2}}\right]=0,
\end{split}
\end{equation}
\begin{equation}\label{3.132}
\begin{split}
&f\left[b_t+\mu_2 c_t+\lambda(b_x+\mu_2 c_x)u_0^2\pm\frac{\mu_2(a-c)}{\sqrt{1+\mu_2^2}}(C_1+\lambda \eta_2 u_0^2)\mp\frac{2b}{\sqrt{1+\mu_2^2}}(C_1+\lambda\eta_2 u_0^2)\right]\\
&+(\eta_2c_t-C_1c_x)+\phi_{1}\left[-b_x-\mu_2c_x\mp\frac{\mu_2\eta_2(a-c)}{\sqrt{1+\mu_2^2}}\pm\frac{2b\eta_2}{\sqrt{1+\mu_2^2}}\right]=0.
\end{split}
\end{equation}
Differentiating (\ref{3.131}) and (\ref{3.132}) with respect to $u_2$ leads to
\begin{equation}\label{3.133}
a_t+\mu_2 b_t+\lambda(a_x+\mu_2 b_x)u_0^2\mp\frac{1}{\sqrt{1+\mu_2^2}}(C_1+\lambda \eta_2 u_0^2)(2\mu_2b+a-c)=0,
\end{equation}
\begin{equation}\label{3.134}
b_t+\mu_2 c_t+\lambda(b_x+\mu_2 c_x)u_0^2\pm\frac{1}{\sqrt{1+\mu_2^2}}(C_1+\lambda \eta_2 u_0^2)(\mu_2a-\mu_2c-2b)=0.
\end{equation}
Taking the second $u_0$ derivative and substituting back gives
\begin{equation}\label{3.135}
a_t+\mu_2 b_t\mp\frac{C_1}{\sqrt{1+\mu_2^2}}(2\mu_2b+a-c)=0,\quad b_t+\mu_2 c_t\pm\frac{C_1}{\sqrt{1+\mu_2^2}}(\mu_2a-\mu_2c-2b)=0,
\end{equation}
\begin{equation}\label{3.136}
\lambda\left[a_x+\mu_2 b_x\mp\frac{\eta_2}{\sqrt{1+\mu_2^2}}(2\mu_2b+a-c)\right]=0, \quad
 \lambda\left[b_x+\mu_2 c_x\pm\frac{\eta_2}{\sqrt{1+\mu_2^2}}(\mu_2a-\mu_2c-2b)\right]=0.
\end{equation}
Substituting (\ref{3.135}) and (\ref{3.136}) into (\ref{3.131}) and (\ref{3.132}) returns
\begin{equation}\label{3.137}
\eta_2b_t-C_1b_x+\phi_{1}\left[-a_x-\mu_2b_x\pm\frac{\eta_2}{\sqrt{1+\mu_2^2}}(2\mu_2b+a-c)\right]=0,
\end{equation}
\begin{equation}\label{3.138}
\eta_2c_t-C_1c_x+\phi_{1}\left[-b_x-\mu_2c_x\mp\frac{\eta_2}{\sqrt{1+\mu_2^2}}(\mu_2a-\mu_2c-2b)\right]=0.
\end{equation}
Linear combinations of  (\ref{3.135}) and (\ref{3.136}) yields
\begin{equation}\label{3.139}
\pm\frac{1}{\sqrt{1+\mu_2^2}}(2\mu_2b+a-c)=\frac{1}{(\lambda\eta_2)^2+C_1^2}[C_1(a_t+\mu_2 b_t)+\lambda^2\eta_2(a_x+\mu_2b_x)],
\end{equation}
\begin{equation}\label{3.140}
\pm\frac{1}{\sqrt{1+\mu_2^2}}(\mu_2a-\mu_2c-2b)=-\frac{1}{(\lambda\eta_2)^2+C_1^2}[C_1(b_t+\mu_2 c_t)+\lambda^2\eta_2(b_x+\mu_2c_x)],
\end{equation}
By substituting these into (\ref{3.137}) and (\ref{3.138}), we get
\begin{equation}\label{3.141}
(\mu_2C_1\phi_{1}+(\lambda\eta_2)^2+C_1^2)(\eta_2b_t-C_1b_x)+C_1\phi_{1}(\eta_2a_t-C_1a_x)=0,
\end{equation}
\begin{equation}\label{3.142}
(\mu_2C_1\phi_{1}+(\lambda\eta_2)^2+C_1^2)(\eta_2c_t-C_1c_x)+C_1\phi_{1}(\eta_2b_t-C_1b_x)=0.
\end{equation}
The differentiation of the Gauss equation (\ref{2.14}) with respect to $t$ and $x$ implies
\begin{equation}\label{3.143}
(\eta_2a_t-C_1a_x)c+(\eta_2c_t-C_1c_x)a-2b(\eta_2b_t-C_1b_x)=0.
\end{equation}
We show that $\eta_2a_t-C_1a_x=\eta_2b_t-C_1b_x=\eta_2c_t-C_1c_x=0$ through the following case analysis.

If $C_1\phi_{1}=0$, from (\ref{3.141}) and (\ref{3.142}), we immediately have $\eta_2b_t-C_1b_x=0$ and $\eta_2c_t-C_1c_x=0$. Substitution into (\ref{3.143}) leads to $\eta_2a_t-C_1a_x=0$.

If $C_1\phi_{1}\neq0$, substituting (\ref{3.141}) and (\ref{3.142}) into (\ref{3.143}), we get
\begin{equation}\label{3.144}
(\eta_2c_t-C_1c_x)[a+Y^2c+2Yb]=0, \quad Y=\frac{\mu_2C_1\phi_{1}+(\lambda\eta_2)^2+C_1^2}{C_1\phi_{1}}.
\end{equation}
Assume $\eta_2c_t-C_1c_x\neq 0$, the Gauss equation (\ref{2.14}) implies
\begin{equation}\label{3.145}
a=Y^2c\mp2Y,\quad b=\pm1-Yc.
\end{equation}
Since $a\neq0$, then $Y$ is a nonzero constant. Substituting (\ref{3.145}) into (\ref{3.135}) leads to
\begin{equation}\label{3.146}
\begin{split}
&(Y-\mu_2)Yc_t\mp\frac{C_1}{\sqrt{1+\mu_2^2}}[2\mu_2^2(\pm1-Yc)+Y^2c\mp2Y-c]=0,
\\&-(Y-\mu_2)c_t\pm\frac{C_1}{\sqrt{1+\mu_2^2}}[-2(\pm1-Yc)+\mu_2(Y^2c\mp2Y-c)]=0.
\end{split}
\end{equation}
Combining the above two equations, we conclude $c=\pm2\mu_2/(1+\mu_2Y)$, namely, $c$ is a nonzero constant. In views of (\ref{3.145}), $a$ and $b$ are constants, reducing (\ref{3.135}) to
\begin{equation}\label{3.147}
2\mu_2b+a-c=0,\quad \mu_2a-\mu_2c-2b=0.
\end{equation}
This implies $b=0$ and $a=c$, which contradicts the Gauss equation (\ref{2.14}). Therefore, we must have $\eta_2a_t-C_1a_x=\eta_2b_t-C_1b_x=\eta_2c_t-C_1c_x=0$. Moreover, we can assume
\begin{equation}\label{3.148}
a=\Phi_1,\quad b=\Phi_2,\quad c=\Phi_3,
\end{equation}
where $\Phi_i=\Phi_i (\xi),\xi=\eta_2 x+C_1t,i=1,2,3$ are differentiable functions satisfying $\Phi_1 \Phi_3\neq0$. Substituting into (\ref{3.135}) and (\ref{3.136}) we get, respectively,
\begin{equation}\label{3.149}
\begin{split}
&C_1(\Phi_{1,\xi}+\mu_2\Phi_{2,\xi})\mp\frac{C_1}{\sqrt{1+\mu_2^2}}(2\mu_2\Phi_2+\Phi_1-\Phi_3)=0,\\
&C_1(\Phi_{2,\xi}+\mu_2\Phi_{3,\xi})\pm\frac{C_1}{\sqrt{1+\mu_2^2}}(\mu_2\Phi_1-\mu_2\Phi_3-2\Phi_2)=0,
\end{split}
\end{equation}
and
\begin{equation}\label{3.150}
\begin{split}
&\lambda\left[\eta_2(\Phi_{1,\xi}+\mu_2\Phi_{2,\xi})\mp\frac{\eta_2}{\sqrt{1+\mu_2^2}}(2\mu_2\Phi_2+\Phi_1-\Phi_3)\right]=0,\\
&\lambda\left[\eta_2(\Phi_{2,\xi}+\mu_2\Phi_{3,\xi})\pm\frac{\eta_2}{\sqrt{1+\mu_2^2}}(\mu_2\Phi_1-\mu_2\Phi_3-2\Phi_2)\right]=0.
\end{split}
\end{equation}
Since $(\lambda\eta_2)^2+C_1^2\neq0$, (\ref{3.149}) and (\ref{3.150}) reduce to
\begin{equation}\label{3.151}
\begin{split}
&\Phi_{1,\xi}+\mu_2\Phi_{2,\xi}\mp\frac{1}{\sqrt{1+\mu_2^2}}(2\mu_2\Phi_2+\Phi_1-\Phi_3)=0,\\
&\Phi_{2,\xi}+\mu_2\Phi_{3,\xi}\pm\frac{1}{\sqrt{1+\mu_2^2}}(\mu_2\Phi_1-\mu_2\Phi_3-2\Phi_2)=0.
\end{split}
\end{equation}
Elimination of $\Phi_3$ in (\ref{3.151}) gives
\begin{equation}\label{3.152}
\mu_2\left(\mp\sqrt{1+\mu_2^2}\Phi_{1,\xi\xi}+2\Phi_{1,\xi}\right)+\mu_2^2\left(\mp\sqrt{1+\mu_2^2}\Phi_{2,\xi\xi}+2\Phi_{2,\xi}\right)
+(1+\mu_2^2)\left(\Phi_{2,\xi}\mp\frac{2}{\sqrt{1+\mu_2^2}}\Phi_2\right)=0,
\end{equation}
that is
\begin{equation}\label{3.153}
\mu_2\Phi_{1,\xi}=\pm\sqrt{1+\mu_2^2}\Phi_2-\mu_2^2\Phi_{2,\xi}\pm\frac{\beta}{\sqrt{1+\mu_2^2}}e^{\pm\frac{2\xi}{\sqrt{1+\mu_2^2}}},
\end{equation}
where $\beta$ is an integration constant constant.

When $\mu_2=0$, it follows from (\ref{3.151}) and (\ref{3.153}) that
\begin{equation}\label{3.154}
b=-\beta e^{\pm2\xi},\xi=\eta_2x+C_1t,\quad c=a\mp a_\xi.
\end{equation}
Similarly, by using the Gauss equation (\ref{2.14}), we get a second order differentiable equation
\begin{equation}\label{3.155}
a^2\mp aa_\xi-\beta^2 e^{\pm4\xi}+1=0,
\end{equation}
whose solution $a$ is defined by (\ref{3.115}). This concludes (ii).

When $\mu_2\neq0$, from (\ref{3.151}) and (\ref{3.153}), we have
\begin{equation}\label{3.156}
\Phi_3=\Phi_1+\Phi,\quad \Phi=\frac{\mu_2^2-1}{\mu_2}\Phi_2-\frac{\beta}{\mu_2}e^{\pm\frac{2\xi}{\sqrt{1+\mu_2^2}}},\xi=\eta_2x+C_1t.
\end{equation}
Substituting into the Gauss equation (\ref{2.14}) leads to $\Phi_1^2+\Phi\Phi_1-\Phi_2^2+1=0$, and thus we find
\begin{equation}\label{3.157}
a=\Phi_1=\frac{-\Phi\pm\sqrt{\Delta}}{2},\quad \Delta=\Phi^2-4(1-\Phi_2^2)>0.
\end{equation}
In views of (\ref{3.156}), we obtain $c$ as in (\ref{3.117}). And then (\ref{3.153}) becomes
\begin{equation}\label{3.158}
\Phi_{2,\xi}[(\mu_2^2+1)\sqrt{\Delta}\pm(\mu_2^2-1)\Phi\pm4\mu_2\Phi_2]\mp2\sqrt{1+\mu_2^2}\sqrt{\Delta}\Phi_2-\frac{2\beta}{\sqrt{1+\mu_2^2}}\Phi e^{\pm\frac{2\xi}{\sqrt{1+\mu_2^2}}}=0.
\end{equation}
Assume the coefficient of $\Phi_{2,\xi}$ in (\ref{3.158}) vanishes, we drive
\begin{align}
&(\mu_2^2+1)\sqrt{\Delta}\pm(\mu_2^2-1)\Phi\pm4\mu_2\Phi_2=0,\label{3.159}
\\&\mp2\sqrt{1+\mu_2^2}\sqrt{\Delta}\Phi_2-\frac{2\beta}{\sqrt{1+\mu_2^2}}\Phi e^{\pm\frac{2\xi}{\sqrt{1+\mu_2^2}}}=0.\label{3.160}
\end{align}
Combining the above two equations requires that $\Phi^2+4\Phi_2^2=0$, which leads to $\Phi_3=\Phi_1$. This contradicts the Gauss equation (\ref{2.14}), proving the coefficient of $\Phi_{2,\xi}$ in (\ref{3.158}) does not vanish. Therefore, we can assume $b_\xi=\Phi_{2,\xi}=g(\xi,b)$, $\xi=\eta_2x+C_1t$, where $g(\xi,b)$ is a differentiable function defined as
\begin{equation}\label{3.161}
g(\xi,b)=\frac{\pm2\sqrt{1+\mu_2^2}b\sqrt{\Delta}+\frac{2\beta}{\sqrt{1+\mu_2^2}}\Phi e^{\pm\frac{2\xi}{\sqrt{1+\mu_2^2}}}}{(\mu_2^2+1)\sqrt{\Delta}\pm(\mu_2^2-1)\Phi\pm4\mu_2 b}.
\end{equation}

Consider a fixed point $\xi_0$ and the associated  initial value problem
\begin{equation}\label{3.162}
\begin{cases}
b_\xi=g(\xi,b),\\
 b(\xi_0)=b_0,
 \end{cases}
\end{equation}
where $g(\xi,b)$ is given by (\ref{3.161}). The function $g(\xi,b)$ and its partial derivative $\partial_b g(\xi,b)$ are both continuous in an open rectangular domain
\begin{equation}\label{3.163}
R=\{(\xi,b):\xi_1<\xi<\xi_2,b_1<b<b_2\}
\end{equation}
containing the initial point $(\xi_0,b_0)$. By the fundamental existence and uniqueness theorem for ordinary differential equations, there exists unique solution $b(\xi)$ on the interval $I=[b_0-\epsilon,b_0+\epsilon]$, where $\epsilon$ is a positive number. Additionally, the interval boundaries $\xi_1$ and $\xi_2$ must be chosen such that $\xi_1<\xi<\xi_2$ intersects with the domain of the solution of (\ref{2.34}). Finally, substituting the expression for $\Phi$ into (\ref{3.158}) produces (\ref{3.128}). This completes (iii).

The converse statement follows directly through straightforward computation.
$\hfill\square$
\end{Proof}

\begin{Proposition}\label{proposition3.8}
Consider a partial differential equation of the form (\ref{2.21}) which describes pseudospherical surfaces, with $f_{ij}$ given by Theorems 2.5 (i). There is no local isometric immersion of the pseudospherical surface, determined by a solution $u(x,t)$ of the equation (\ref{2.21}), for which the coefficients $a$, $b$, $c$ of the second fundamental form depend on $x$, $t$, $u_0, \dots, u_l$, $w_1, \dots, w_m$, $v_1,\dots, v_n$, $1 \leq l,m,n < \infty$.
\end{Proposition}

\begin{Proof}
Suppose $f_{21} =0$, which implies $\mu_2=\eta_2=0$. Consequently, Lemma 3.2 holds and (\ref{2.15}) and (\ref{2.16}) become
\begin{equation}\label{3.164}
f_{11}[a_t+\lambda u_0^2 a_x-2b(\phi_{3}\mp\phi_{1})\mp(a-c)\phi_{2}]+\phi_{1}\left(a_x\pm2b\frac{\theta}{\nu}\right)-\phi_{2}\left[b_x\pm(a-c)\frac{\theta}{\nu}\right]=0,
\end{equation}
\begin{equation}\label{3.165}
f_{11}[b_t+\lambda u_0^2 b_x+(a-c)(\phi_{3}\mp\phi_{1})\mp2b\phi_{2}]+\phi_{1}\left[b_x\mp(a-c)\frac{\theta}{\nu}\right]-\phi_{2}\left(c_x\pm2b\frac{\theta}{\nu}\right)=0,
\end{equation}
Differentiating (\ref{3.164}) and (\ref{3.165}) with respect to $u_2$, we have
\begin{equation}\label{3.166}
a_t+\lambda u_0^2 a_x-2b(\phi_{3}\mp\phi_{1})\mp(a-c)\phi_{2}=0,
\end{equation}
\begin{equation}\label{3.167}
b_t+\lambda u_0^2 b_x+(a-c)(\phi_{3}\mp\phi_{1})\mp2b\phi_{2}=0,
\end{equation}
where $\phi_{3}\mp\phi_{1}=\mp\left(2\lambda/\theta-\theta C_2 e^{\theta u_0}+2\lambda u_0\right)/\nu$ and $\phi_{2}=\pm\left(2\lambda/\theta-\theta C_2 e^{\theta u_0}+2\lambda u_0\right)u_1$. Taking the $u_1$ derivative of (\ref{3.166}) and (\ref{3.167}) produces
\begin{equation}\label{3.168}
-(a-c)\left(\frac{2\lambda}{\theta}-\theta C_2 e^{\theta u_0}+2\lambda u_0\right)=0,
\end{equation}
\begin{equation}\label{3.169}
-2b\left(\frac{2\lambda}{\theta}-\theta C_2 e^{\theta u_0}+2\lambda u_0\right)=0,
\end{equation}
since $\lambda^2+C_2^2\neq 0$ and $\theta \neq 0$, then $2\lambda/\theta-\theta C_2 e^{\theta u_0}+2\lambda u_0\neq 0$. Hence, (\ref{3.168}) and (\ref{3.169}) return $a-c=b=0$, which contradicts the Gauss equation (\ref{2.14}).

Suppose $f_{21} \neq0$, from Lemma 3.4,  the equations (\ref{2.14})-(\ref{2.16}) form an inconsistent system. Hence, Lemma 3.3 holds and (\ref{2.15}) and (\ref{2.16}) become
\begin{equation}\label{3.170}
\begin{split}
&f_{11}\left[a_t+\mu_2 b_t+\lambda (a_x+\mu_2 b_x)u_0^2-2b\left(\phi_{3}\mp\sqrt{1+\mu_2^2}\phi_{1}\right)
 +(a-c)\left(\mu_2\phi_{3}\mp\sqrt{1+\mu_2^2}\phi_{2}\right)\right]
\\& -\phi_{1}\left(a_x\mp2b\frac{\theta+\nu\mu_2\eta_2}{\nu\sqrt{1+\mu_2^2}}\right)
-\phi_{2}\left[b_x\pm(a-c)\frac{\theta+\nu\mu_2\eta_2}{\nu\sqrt{1+\mu_2^2}}\right]+\eta_2(b_t+\lambda u_0^2 b_x)+\eta_2 \phi_{3}(a-c)=0,
\end{split}
\end{equation}
\begin{equation}\label{3.171}
\begin{split}
&f_{11}\left[b_t+\mu_2 c_t+\lambda (b_x+\mu_2 c_x)u_0^2+(a-c)\left(\phi_{3}\mp\sqrt{1+\mu_2^2}\phi_{1}\right)
+2b\left(\mu_2\phi_{3}\mp\sqrt{1+\mu_2^2}\phi_{2}\right)\right]
\\&-\phi_{1}\left(b_x\pm(a-c)\frac{\theta+\nu\mu_2\eta_2}{\nu\sqrt{1+\mu_2^2}}\right)
-\phi_{2}\left[c_x\pm2b\frac{\theta+\nu\mu_2\eta_2}{\nu\sqrt{1+\mu_2^2}}\right]+\eta_2(c_t+\lambda u_0^2 c_x)+2b\eta_2 \phi_{3}=0,
\end{split}
\end{equation}
where
\begin{equation}\label{3.172}
\begin{split}
&\phi_{3}\mp\sqrt{1+\mu_2^2}\phi_{1}=-\left(\frac{2\lambda}{\theta}-\theta C_2 e^{\theta u_0}+2\lambda u_0\right)(\mu_2 u_1-\frac{\eta_3}{\theta}), \\ &\mu_2\phi_{3}\mp\sqrt{1+\mu_2^2}\phi_{2}=-\left(\frac{2\lambda}{\theta}-\theta C_2 e^{\theta u_0}+2\lambda u_0\right)\left(u_1+\frac{\mu_2\eta_3-\mu_3\eta_2}{\theta}\right).
\end{split}
\end{equation}
Differentiating (\ref{3.170}) and (\ref{3.171}) with respect to $u_2$ gives
\begin{equation}\label{3.173}
a_t+\mu_2 b_t+\lambda (a_x+\mu_2 b_x)u_0^2-2b\left(\phi_{3}\mp\sqrt{1+\mu_2^2}\phi_{1}\right)+(a-c)\left(\mu_2\phi_{3}\mp\sqrt{1+\mu_2^2}\phi_{2}\right)=0,
\end{equation}
\begin{equation}\label{3.174}
b_t+\mu_2 c_t+\lambda (b_x+\mu_2 c_x)u_0^2+(a-c)\left(\phi_{3}\mp\sqrt{1+\mu_2^2}\phi_{1}\right)+2b\left(\mu_2\phi_{3}\mp\sqrt{1+\mu_2^2}\phi_{2}\right)=0.
\end{equation}
Taking the $u_1$ derivative of (\ref{3.173}) and (\ref{3.174}), we have
\begin{equation}\label{3.175}
-2b\left(\phi_{3}\mp\sqrt{1+\mu_2^2}\phi_{1}\right)_{u_1}+(a-c)\left(\mu_2\phi_{3}\mp\sqrt{1+\mu_2^2}\phi_{2}\right)_{u_1}=0,
\end{equation}
\begin{equation}\label{3.176}
(a-c)\left(\phi_{3}\mp\sqrt{1+\mu_2^2}\phi_{1}\right)_{u_1}+2b\left(\mu_2\phi_{3}\mp\sqrt{1+\mu_2^2}\phi_{2}\right)_{u_1}=0.
\end{equation}
Combining the above two equation returns
\begin{equation}\label{3.177}
\left[(a-c)^2+4b^2\right]\left(\mu_2\phi_{3}\mp\sqrt{1+\mu_2^2}\phi_{2}\right)_{u_1}=0.
\end{equation}
Since $ac-b^2=-1$, we drive $(a-c)^2+b^2\neq0$. Thus, above equation holds if and only if $\lambda=C_2=0$, which is a contradiction.
$\hfill\square$
\end{Proof}

\begin{Proposition}\label{proposition3.9}
Consider a partial differential equation of the form (\ref{2.21}) which describes pseudospherical surfaces, with $f_{ij}$ given by Theorems 2.5 (ii). There is no local isometric immersion of the pseudospherical surface, determined by a solution $u(x,t)$ of the equation (\ref{2.21}), for which the coefficients $a$, $b$, $c$ of the second fundamental form depend on $x$, $t$, $u_0, \dots, u_l$, $w_1, \dots, w_m$, $v_1,\dots, v_n$, $1 \leq l,m,n < \infty$.
\end{Proposition}

\begin{Proof}
The non-vanishing condition $\eta_2\neq0$ guarantees that $f_{21}\neq  0$ on an open set. Lemma 3.4 shows that the system of equations (\ref{2.14})-(\ref{2.16}) is inconsistent. Consequently, Lemma 3.3 holds and (\ref{2.15}) and (\ref{2.16}) become
\begin{equation}\label{3.178}
\begin{split}
&f_{11}[a_t+\mu_2 b_t+\lambda(a_x+\mu_2 b_x)u_0^2\mp2b\eta_3\tau\varphi e^{\pm\tau u_1}\pm(a-c)(\mu_2\eta_3-\mu_3\eta_2)\tau\varphi e^{\pm\tau u_1}]\\
&+\eta_2(b_t\mp\tau\varphi e^{\pm\tau u_1}b_x)-\phi_{1}[a_x+\mu_2 b_x-2b\eta_3+(a-c)(\mu_2\eta_3-\mu_3\eta_2)]=0,
\end{split}
\end{equation}
\begin{equation}\label{3.179}
\begin{split}
&f_{11}[b_t+\mu_2 c_t+\lambda(b_x+\mu_2 c_x)u_0^2\pm(a-c)\eta_3\tau\varphi e^{\pm\tau u_1}\pm2b(\mu_2\eta_3-\mu_3\eta_2)\tau\varphi e^{\pm\tau u_1}]\\
&+\eta_2(c_t\mp\tau\varphi e^{\pm\tau u_1}c_x)-\phi_{1}[b_x+\mu_2 c_x+(a-c)\eta_3+2b(\mu_2\eta_3-\mu_3\eta_2)]=0.
\end{split}
\end{equation}
Differentiating (\ref{3.178}) and (\ref{3.179}) with respect to $u_2$, we see
\begin{equation}\label{3.180}
a_t+\mu_2 b_t+\lambda(a_x+\mu_2 b_x)u_0^2\mp2b\eta_3\tau\varphi e^{\pm\tau u_1}\pm(a-c)(\mu_2\eta_3-\mu_3\eta_2)\tau\varphi e^{\pm\tau u_1}=0,
\end{equation}
\begin{equation}\label{3.181}
b_t+\mu_2 c_t+\lambda(b_x+\mu_2 c_x)u_0^2\pm(a-c)\eta_3\tau\varphi e^{\pm\tau u_1}\pm2b(\mu_2\eta_3-\mu_3\eta_2)\tau\varphi e^{\pm\tau u_1}=0.
\end{equation}
Taking the $u_1$ derivative of (\ref{3.180}) and (\ref{3.181}) leads to
\begin{equation}\label{3.182}
-2b\eta_3+(a-c)(\mu_2\eta_3-\mu_3\eta_2)=0,
\end{equation}
\begin{equation}\label{3.183}
(a-c)\eta_3+2b(\mu_2\eta_3-\mu_3\eta_2)=0,
\end{equation}
where the conditions $\tau>0$ and $\varphi\neq0$ have been used. Since $(a-c)^2+b^2\neq0$, it follows from the above system that $\eta_3=\mu_2\eta_3-\mu_3\eta_2=0$. This leads to
\begin{equation}\label{3.184}
\Delta_{13}=\eta_3[\pm\tau\varphi e^{\pm\tau u_1}f_{11}-\phi_1]=0,\quad \Delta_{23}=(\mu_2\eta_3-\mu_3\eta_2)[\pm\tau\varphi e^{\pm\tau u_1}f_{11}-\phi_1]=0,
\end{equation}
which contradicts the condition (\ref{2.18}).
$\hfill\square$
\end{Proof}

Finally, the comprehensive proof of Theorem 1.1 follows from Lemma 3.1-3.4 and Propositions 3.5-3.9. And the proof of Corollary 1.2 follows from Proposition 3.7.
$\hfill\square$

\section*{Acknowledgements} Guo's research was supported by Northwest University Graduate Research and Innovation Program CX2024133.


\begin{thebibliography}{99}

\bibitem{Chern1986}
Chern S.S., Tenenblat K., Pseudospherical surfaces and evolution equations, \emph{Stud. Appl. Math.} 74 (1986), 55-83.

\bibitem{Sasaki1979}
Sasaki R., Soliton equations and pseudospherical surfaces, \emph{Nuclear Phys. B} 154 (1979), 343-357.

\bibitem{Jorge1987}
Jorge L., Tenenblat K., Linear problems associated with evolution equations of the type $u_{tt} = F(u, u_x, u_{xx}, u_t)$, \emph{Stud. Appl. Math.} 77 (1987), 103-117.

\bibitem{Rabelo1989}
Rabelo M.L., On equations which describe pseudospherical surfaces,  \emph{Stud. Appl. Math.} 81 (1989), 221-248.

\bibitem{Reyes1998}
Reyes E.G., Pseudospherical surfaces and integrability of evolution equations, \emph{J. Differ. Equ.} 147 (1998), 195-230.

\bibitem{Ding2002}
Ding Q., Tenenblat K., On differential systems describing surfaces of constant curvature, \emph{J. Differ. Equ.} 184 (2002), 185-214.

\bibitem{Silva2015}
Silva T.C., Tenenblat K., Third order differential equations describing pseudospherical surfaces, \emph{J. Differ. Equ.} 259 (2015), 4897-4923.

\bibitem{Neto2010}
Neto V.G., Fifth order evolution equations describing pseudospherical surfaces, \emph{J. Differ. Equ.} 249 (2010), 2822-2865.

\bibitem{Ferraioli2020}
Ferraioli D.C., Silva T.C., Tenenblat K., A class of quasilinear second order partial differential equations which describe spherical or pseudospherical surfaces, \emph{J. Differ. Equ.} 268 (2020), 7164-7182.

\bibitem{Kelmer2022}
Kelmer F., Tenenblat K., On a class of systems of hyperbolic equations describing pseudospherical or spherical surfaces, \emph{J. Differ. Equ.} 339 (2022), 372-394.

\bibitem{Ferraioli2024}
Ferraioli D.C., Silva T.C., A class of third order quasilinear partial differential equations describing spherical or pseudospherical surfaces, \emph{J. Differ. Equ.} 379 (2024), 524-568.

\bibitem{Kelmer2025}
Kelmer F., Tenenblat K., Systems of differential equations of higher order describing pseudo-spherical or spherical surfaces, \emph{J. Differ. Equ.} 424 (2025), 833-858.

\bibitem{Kahouadji2015}
Kahouadji N., Kamran N., Tenenblat K., Local Isometric Immersions of Pseudo-Spherical Surfaces and Evolution Equations, In: Guyenne P., Nicholls D., Sulem C. (eds) Hamiltonian Partial Differential Equations and Applications, Fields Institute Communications, vol 75, Springer, New York, NY, 2015.

 \bibitem{Kahouadji2016}
Kahouadji N., Kamran N., Tenenblat K., Second-order equations and local isometric immersions of pseudo-spherical surfaces, \emph{Commun. Anal. Geom.} 24 (3) (2016), 605-643.

\bibitem{Silva2016}
Castro Silva T., Kamran N., Third-order differential equations and local isometric immersions of pseudospherical surfaces, \emph{Commun. Contemp. Math.} 18 (6) (2016), 1650021.

\bibitem{Ferraioli2017}
Ferraioli D.C., de Oliveira L.A., Local isometric immersions of pseudospherical surfaces described by evolution equations in conservation law form, \emph{J. Math. Anal. Appl.} 446 (2017), 1606-1631.

\bibitem{Kahouadji2019}
Kahouadji N., Kamran N., Tenenblat K., Local isometric immersions of pseudo-spherical surfaces and k-th order evolution equations, \emph{Commun. Contemp. Math.} 21 (04) (2019), 1850025.

\bibitem{Ferraioli2022}
Ferraioli D. C., Silva T. C., Tenenblat K., Isometric immersions and differential equations describing pseudospherical surfaces, \emph{J. Math. Anal. Appl.} 511 (2022), 126091.

\bibitem{Rabelo1990}
Rabelo M. L., Tenenblat K., On equations of type $u_{xt}=F(u,u_x)$ which describe pseudospherical surfaces, \emph{J.Math. Phys.} 6 (1990), 1400-1407.

\bibitem{Ferraioli2016}
Ferraioli D. C., de Oliveira Silva L., Second order evolution equations which describe pseudospherical surfaces, \emph{J. Differ. Equ.} 260 (2016), 8072-8108.

\bibitem{Bour1862}
Bour E., Th$\acute{e}$orie de la d$\acute{e}$formation des surfaces, \emph{J. $ \acute{E}col$. Imper. Polytech.} 19 (39) (1862), 1-48.

\bibitem{Freire2022}
Freire I. L., Tito R. S., A Novikov equation describing pseudospherical surfaces, its pseudo-potentials, and local isometric immersions, \emph{Stud. Appl. Math.}, 148 (2022), 758-772.

\bibitem{Novikov2009}
Novikov V., Generalizations of the Camassa-Holm equation, \emph{J. Phys. A Math. Theor.} 42 (2009), 342002.

 \bibitem{Cartan1970}
Cartan H., Formes diff$\acute{e}$rentielles, Hermann, Paris, 1970.

\bibitem{Ivey2003}
Ivey T.A., Landsberg J.M., Cartan for Beginners: Differential Geometry via Moving Frames and Exterior Differential Systems, AMS, 2003.

\bibitem{Tenenblat1998}
Tenenblat K., Transformations of Manifolds and Applications to Differential Equations, Addison-Wesley/Longman, London, UK, 1998.

\bibitem{Clelland2017}
Clelland J. N., From Frenet to Cartan: the method of moving frames, AMS, 2017.

\end{thebibliography}
\end{document}